\documentclass[aps,prd,floatfix,nofootinbib,12pt]{revtex4}
\usepackage{epsfig}
\usepackage{bm}
\usepackage{dcolumn}
\usepackage{latexsym}

\begin{document}

\title{Applying Complex Langevin Simulations to Lattice QCD 
at Finite Density}

\author{J.~B.~Kogut}
\affiliation{Department of Energy, Division of High Energy Physics, Washington,
DC 20585, USA \\
and \\
Department of Physics -- TQHN, University of Maryland, 82 Regents Drive,
College Park, MD 20742, USA}
\author{D.~K.~Sinclair}
\affiliation{HEP Division, Argonne National Laboratory, 9700 South Cass Avenue,
Lemont, IL 60439, USA}

\begin{abstract}
We study the use of the complex-Langevin equation (CLE) to simulate lattice QCD
at a finite chemical potential ($\mu$) for quark-number, which has a complex
fermion determinant that prevents the use of standard simulation methods
based on importance sampling. Recent enhancements to the CLE specific to lattice
QCD inhibit runaway solutions which had foiled earlier attempts to use it for
such simulations. However, it is not guaranteed to produce correct results.
Our goal is to determine under what conditions the CLE yields correct values
for the observables of interest. Zero temperature simulations indicate that
for moderate couplings, good agreement with expected results is obtained for
small $\mu$ and for $\mu$ large enough to reach saturation, and that this
agreement improves as we go to weaker coupling. For intermediate $\mu$ values
these simulations do not produce the correct physics. We compare our results
with those of the phase-quenched approximation. Since there are indications
that correct results might be obtained if the CLE trajectories remain close to
the $SU(3)$ manifold, we study how the distance from this manifold depends on
the quark mass and on the coupling. We find that this distance decreases with
decreasing quark mass and as the coupling decreases, i.e. as the simulations
approach the continuum limit.
\end{abstract}

\maketitle

\section{Introduction}

QCD at finite baryon/quark-number density describes nuclear matter, the 
constituent of neutron stars and the interiors of heavy nuclei. We are
interested in calculating the phase diagram for nuclear matter. Knowing the
properties of nuclear matter can yield an equation-of-state and a better
description of neutron stars. In addition it can yield information on the
interaction of nuclear matter with particles passing through it. Nuclear matter
at high temperatures undergoes a transition to a quark-gluon plasma. Such
transitions can be observed in relativistic heavy-ion collisions. While for
low densities this transition is expected to be a crossover, it is believed
that for high densities it becomes a first-order transition. The critical
end-point is the second-order transition (critical point) where this change
occurs.

Finite density QCD has a sign problem which prevents the direct application of
standard lattice QCD simulation methods, which rely on importance sampling. When
finite density is implemented by use of a quark-number chemical potential,
$\mu$, the sign problem manifests itself by making the fermion determinant 
complex, with a real part of indefinite sign. Simulations using the complex
Langevin equation (CLE) 
\cite{Parisi:1984cs,Klauder:1983nn,Klauder:1983zm,Klauder:1983sp}
can accommodate such complex actions. However, the CLE can only be shown to
yield correct values for observables if the space over which the fields evolve
is compact, the drift (force) term is holomorphic in the fields, and the
solutions are ergodic
\cite{Aarts:2009uq,Aarts:2011ax,Nagata:2015uga,Nishimura:2015pba,Nagata:2016vkn,
Aarts:2017vrv,Seiler:2017wvd,Aarts:2017hqp,Nagata:2018net}.

For QCD, implementation of the CLE requires extending the gauge-field manifold
from $SU(3)$ to $SL(3,C)$. Keeping the action holomorphic in the fields except
on a space of measure zero requires making the action a function of the gauge
fields and their inverses only. Then the drift term is meromorphic in the
gauge fields, having poles at the zeros of the fermion determinant. Early
attempts at simulations were frustrated by runaway behaviour which could not
be controlled by adaptively decreasing the updating interval. Recently it was 
discovered that, for small enough couplings, this behaviour could be tamed by
`gauge-cooling', gauge transforming the fields after each update to keep them
as near as possible to the $SU(3)$ manifold \cite{Seiler:2012wz}.
When this is done, the gauge fields appear to evolve over a compact manifold,
at least for weak enough coupling. It remains to be determined for what if any
range of quark masses, lattice spacings, chemical potentials and temperatures,
and for what choices of lattice actions, the CLE produces correct values for
chosen observables, despite the presence of poles in the drift term and/or in 
the operators defining these observables.

Extensive studies have been performed of heavy dense lattice QCD at finite 
$\mu$ using the CLE. 
\cite{Aarts:2008rr,Aarts:2013uxa,Aarts:2014bwa,Aarts:2016qrv,Langelage:2014vpa,
Rindlisbacher:2015pea}.
Some CLE simulations of lattice QCD have been performed at finite $\mu$ with
lighter quark masses, both on small lattices and at finite temperatures
\cite{Sexty:2013ica,Aarts:2014bwa,Fodor:2015doa,Nagata:2016mmh,Tsutsui:2018jva,
Scherzer:2018udt}.
Preliminary work directed towards zero temperatures has been reported
\cite{Ito:2018jpo}. In heavy dense lattice QCD at finite chemical potential,
the zeros of the fermion determinant (poles in the drift term) only affect the
CLE results very close to the transition, provided one excludes the regions
where the real part of the fermion determinant is negative, when necessary.
Alternatively, good results for heavy quarks can be obtained by restricting
the length of CLE trajectories to keep them close to the $SU(3)$ manifold. For
lighter quarks, CLE simulations are found to produce good results at high
temperatures (above the finite temperature phase transition), but the situation
at lower temperatures is less clear.

We simulate zero temperature lattice QCD with 2 flavours (tastes) of staggered
quarks at $\mu$ values from zero to saturation, using the CLE. Here we are
interested in the phase transition from hadronic to nuclear matter which
should occur at $\mu \sim m_N/3$. Random matrix theories (RMT) related to QCD at
finite $\mu$ suggest that when CLE simulations fail, they produce the results
of the phase-quenched model (the theory where the fermion determinant is
replaced by its magnitude), which has a transition to a superfluid state with
a pion-like condensate at $\mu \approx m_\pi/2$. This has been observed by
Mollgard and Splittorff \cite{Mollgaard:2013qra} who simulate the Osborn RMT
\cite{Osborn:2004rf,Bloch:2012bh} and find that the CLE fails for small masses,
approaching phase-quenched results for small-enough masses. They suggest a
solution in a subsequent paper \cite{Mollgaard:2014mga}. Bloch {\it et al.}
\cite{Bloch:2017sex} using the Stephanov RMT \cite{Stephanov:1996ki} 
(which has a non-trivial phase structure) find that the CLE generates 
phase-quenched results. (Note that other random matrix CLE simulations seem
more optimistic \cite{Nagata:2016alq}.) For this reason we also perform RHMC
simulations of the phase-quenched theory at the same values of $\beta=6/g^2$
and quark mass $m$ over the same range of $\mu$ values and on the same lattice
size as the full theory, for comparison. Hence it is important that we choose
$m$ such that $m_N/3$ is significantly larger than $m_\pi/2$. We perform our
simulations at $\beta=5.6$, $m=0.025$ (in lattice units) on a $12^4$ lattice
and at $\beta=5.7$, $m=0.025$ on a $16^4$ lattice. Preliminary results were
reported at Lattice 2015--2018. See \cite{Sinclair:2018rbk} and its references
to our earlier talks.

At $\beta=5.6$, the CLE measurement of the plaquette for $\mu=0$ exhibits a 
systematic error of $\approx 0.31$~\%, while at saturation it shows a systematic
error of  $\approx 1.43$~\%. These should be compared with the increase in 
value of the plaquette over this range which is $\approx 9.2$~\%. At $\beta=5.7$
the CLE plaquette measurement has a systematic error of $\approx 0.16$~\%, while
at saturation the systematic error is $\approx 0.3$~\%. The increase in the
known value of the plaquette over this range is $\approx 6.6$~\%. We note that
for both $\beta$s, the plaquette values at saturation show excellent agreement
with those obtained from CLE simulations of $SU(3)$ lattice gauge theory in 
the absence of quarks, as expected.

At $\beta=5.6$, the CLE measurement of the chiral condensate at $\mu=0$ lies
$\approx 6.9$~\% below the correct value, while at $\beta=5.7$ the chiral
condensate predicted by the CLE is $\approx 1.22$~\% lower than the known
value. For $\mu < m_\pi/2$ there is a similar improvement in the CLE-predicted
chiral condensate between $\beta=5.6$ and $\beta=5.7$. At $\beta=5.7$ and $0.5
\le \mu \le 0.9$ the CLE produces values of the chiral condensate,
quark-number density and plaquette in good agreement with the phase-quenched
theory. However, the Wilson Line/ Polyakov Loop indicates that the fermion
determinant is still complex. Although the CLE results appear to be approaching
the correct physics for very small and large $\mu$ as the coupling decreases
towards the continuum limit, they still fail to produce the expected physics
in the transition region. The transition to nuclear matter appears to start
at $\mu$ even less than $m_\pi/2$ instead of $\mu \approx m_N/3$.

%What we find is that there is reasonable agreement with the expected values for
%observables at $\beta=5.6$, for $\mu << m_\pi/2$ and $\mu >> m_N/3$, and
%significantly better agreement for $\beta=5.7$. However, for $\mu$ in the
%transition region, these simulations fail to produce the expected physics. The
%transition to nuclear matter appears to start at $\mu$ even less than
%$m_\pi/2$ instead of $\sim m_N/3$. There is some indication at $\beta=5.7$
%that the CLE might yield correct results for $\mu \gtrsim m_N/3$.

Since there are indications that the CLE might produce correct results when its
trajectories remain close to the $SU(3)$ manifold, we perform a systematic
study of how the average distance to this manifold (measured using the 
`unitarity norm') depends on the quark mass and the coupling. We find that
this norm decreases with decreasing quark mass and with decreasing coupling,
i.e. as we approach the continuum limit.

In section~2 we present the formulation of the CLE, we use. Section~3 describes
our simulations at $\beta=5.6$ and $\beta=5.7$ and presents results. In 
section~4 we present our simulations to determine how the unitarity norm
depends on quark mass $m$ and lattice coupling $g$. Section~5 gives a summary,
discussion and conclusions.

\section{Complex Langevin for finite density Lattice QCD}

If $S(U)$ is the gauge action after integrating out the quark fields, the
Langevin equation for the evolution of the gauge fields $U$ in Langevin 
time $t$ is:
\begin{equation}
-i \left(\frac{d}{dt}U_l\right)U_l^{-1} = -i \frac{\delta}{\delta U_l}S(U)
+\eta_l
\end{equation}
where $l$ labels the links of the lattice, and 
$\eta_l=\sum_a \eta^a_l\lambda^a$. Here $\lambda_a$ are the Gell-Mann 
matrices for $SU(3)$. $\eta^a_l(t)$ are Gaussian-distributed random 
numbers normalized so that:
\begin{equation}
\langle\eta^a_l(t)\eta^b_{l'}(t')\rangle=\delta^{ab}\delta_{ll'}\delta(t-t')
\end{equation}
We note in passing that the discretized Langevin Equation is the limiting
case of the Hybrid Molecular Dynamics method where each trajectory has only a 
single update.

The complex-Langevin equation has the same form except that the $U$s are now
in $SL(3,C)$. $S$, now $S(U,\mu)$ is 
\begin{equation}
S(U,\mu) = \beta\sum_\Box \left\{1-\frac{1}{6}Tr[UUUU+(UUUU)^{-1}]\right\}
-\frac{N_f}{4}{\rm Tr}\{ln[M(U,\mu)]\}
\end{equation}
where $M(U,\mu)$ is the unimproved staggered Dirac operator with quark-number
chemical potential $\mu$, for a single staggered fermion field (corresponding
to 4 continuum flavours). Note: backward links are represented by $U^{-1}$ not
$U^\dag$. Note also that we have chosen to keep the noise term $\eta$ real.

To simulate the time evolution of the gauge fields we use the partial 
second-order formalism of Fukugita, Oyanagi and Ukawa. 
\cite{Ukawa:1985hr,Fukugita:1986tg,Fukugita:1988qs}
For an update of the fields by a `time' increment $dt$, this gives:
\begin{eqnarray}
U^{(n+1/2)} &=& e^{X_0}U^{(n)} \\ \nonumber
X_0 &=& dt\frac{\delta}{\delta U}S(U^{(n)},\mu)+i\sqrt{dt}\eta^{(n)} \\
                                                          \nonumber
U^{(n+1)}   &=& e^{\gamma(X_0+X_1)}U^{(n)} \\ \nonumber
X_1 &=& dt\frac{\delta}{\delta U}S(U^{(n+1/2)},\mu)+i\sqrt{dt}\eta^{(n)}
\end{eqnarray}
where $\gamma=\frac{1}{2}+\frac{1}{4}dt$ and the Gaussian noise $\eta$ 
is normalized such that:
\begin{equation}
\langle\eta^{a(m)}_l\eta^{b(n)}_{l'}\rangle=
    \left(1-\frac{3}{2}dt\right)\delta^{ab}\delta_{ll'}\delta^{mn}
\end{equation}
To proceed, we replace the spacetime trace with a stochastic estimator $\xi$
\begin{equation}
{\rm Tr}\{ln[M(U,\mu)]\} \rightarrow \xi^\dagger\{ln[M(U,\mu)]\}\xi \;,
\end{equation}
$\xi$ is a vector over space-time and colour of Gaussian random 
numbers, normalized such that:
\begin{equation}
\langle\xi^{*i(m)}(x)\xi^{j(n)}(y)\rangle = \delta^{ij}\delta_{xy}\delta^{mn}
\end{equation}
which means, in particular, that the $\xi$s in $X_0$ and $X_1$ are
independent, unlike the $\eta$s. After performing $\frac{\delta}{\delta U}$ of
$ln(M)$ it is useful to use the cyclic property of the trace to rearrange the
terms proportional to $U$ and $U^{-1}$ prior to introducing the stochastic
estimators, so that this operator is antihermitian when $\mu=0$ and $U$ is
unitary. That way, in this special case, the complex Langevin equation becomes
the real Langevin equation.

We apply adaptive updating. If $f_{ij}(l)$ are the components of the drift
term, we define
\begin{equation}
f_{\rm max}=\left.{\stackrel{\textstyle\rm MAX} {l,i,j}}\right.
            \left|f_{ij}(l)\right|
\end{equation}
where $l$ runs over the links of the lattice. $i=1,2,3$, $j=1,2,3$ are the
colour indices. Then, if $f_{\rm max} > 1$, we replace the input updating 
increment $dt$ by the adaptive increment 
\begin{equation}
dt_{\rm adaptive}=\frac{dt}{f_{\rm max}}
\end{equation} 
for the current update. Because the Dirac operator is often ill-conditioned, we
use 64-bit floating point precision throughout.

After each update, we adaptively gauge fix to the gauge which minimizes the
unitarity norm:
\begin{equation}
F(U) = \frac{1}{4V}\sum_{x,\mu}Tr\left[U^\dag U + (U^\dag U)^{-1} - 2\right]
                                            \ge 0
\end{equation}
which equals $0$, if and only if $U$ is unitary. $V$ is the space-time volume
of the lattice. $F(U)$ is a measure of the lattice averaged distance of the
gauge fields from the $SU(3)$ manifold. We implement gauge cooling after each
updating following the method described in \cite{Seiler:2012wz}, equations
8--10. Here $\epsilon=dt$ the input increment for our CLE updating, and we
choose $\alpha=1/4$. We make this updating adaptive by the following ansatz.
If $G(n)=\sum_{a}\lambda_aG_a(n)$, $G(n)_{norm}=\sqrt{{\rm Tr}(G(n)^2)}$, then
if $\alpha G_{norm}(n) > 1$, we replace $G(n)$ by  $G(n)/[\alpha G_{norm}(n)]$
for this gauge-cooling step.

\section{CLE simulations of lattice QCD at finite $\mu$ and zero temperature}

We simulate two-flavour lattice QCD at $\beta=5.6$, $m=0.025$  on a $12^4$
lattice, and at $\beta=5.7$, $m=0.025$ on a $16^4$ lattice, at 
$0 \le \mu \le 1.5$. Note $\mu=1.5$ is well into the saturation domain, where
all fermion levels are filled and the fermion number density, normalized to 1
staggered fermion field (4 flavours/tastes)is 3. The input $dt$ is chosen to 
be $0.01$. Our runs for individual $\mu$ values vary between $10^6$ and 
$3 \times 10^6$ updates in length. Discarding the first fifth of each run for 
equilibration, this makes the length of each run in Langevin time units vary 
between 80 and over 1000, with the shortest runs being in the saturation regime.
For most of the runs, we choose 5 gauge-cooling steps after each CLE update.
 
To test our choice of $dt$ and the number of gauge-cooling steps per update, as
well as to observe the finite-lattice-size effects we ran a number of test runs
at $\beta=5.6$, $\mu=0$ where we could compare our CLE simulations with those 
performed using the exact RHMC algorithm and with the corresponding real
Langevin equation (RLE). The results of these simulations are summarized in 
tables~\ref{tab:plaquette},\ref{tab:pbp},\ref{tab:unorm}

\begin{table}[htb]
\setlength{\tabcolsep}{10pt}
\begin{tabular}{lllllllll}
\hline
& lattice & $\beta$ & $\mu$ & $dt$ & $dt_{adaptive}$ & cools & start &
plaquette \\
\hline
RHMC & $12^4$ & 5.6 & 0.0 & & & & ordered & 0.43552(5) \\
RHMC & $16^4$ & 5.6 & 0.0 & & & & ordered & 0.43556(2) \\
RLE  & $16^4$ & 5.6 & 0.0 & 0.01 & 0.00103 & 0 & ordered & 0.43566(3) \\
CLE  & $12^4$ & 5.6 & 0.0 & 0.01 & 0.000309 & 5 & ordered & 0.43667(9) \\
CLE  & $12^4$ & 5.6 & 0.0 & 0.01 & 0.000299 & 15 & disordered & 0.43672(12)\\
CLE  & $12^4$ & 5.6 & 0.0 & 0.01 & 0.000294 & 100 & ordered & 0.43686(10)\\
CLE  & $16^4$ & 5.6 & 0.0 & 0.01 & 0.000183 & 5 & ordered & 0.43681(6) \\
CLE  & $16^4$ & 5.6 & 0.0 & 0.005 & 0.000094 & 5 & $dt=0.01$ & 0.43682(7) \\
\hline
\end{tabular}
\caption{Plaquettes from simulations at $\beta=5.6$, $\mu=0$ for various
types of simulation and choice of parameters.}
\label{tab:plaquette}
\end{table}

\begin{table}[htb]                                                      
\setlength{\tabcolsep}{10pt}   
\begin{tabular}{lllllllll}                                                     
\hline                                                                         
& lattice & $\beta$ & $\mu$ & $dt$ & $dt_{adaptive}$ & cools & start &
$\langle\bar{\psi}\psi\rangle$ \\
\hline
RHMC & $12^4$ & 5.6 & 0.0 & & & & ordered & 0.2133(11) \\
RHMC & $16^4$ & 5.6 & 0.0 & & & & ordered & 0.2158(3)  \\
RLE  & $16^4$ & 5.6 & 0.0 & 0.01 & 0.00103 & 0 & ordered & 0.2160(5) \\
CLE  & $12^4$ & 5.6 & 0.0 & 0.01 & 0.000309 & 5 & ordered & 0.1993(15) \\
CLE  & $12^4$ & 5.6 & 0.0 & 0.01 & 0.000299 & 15 & disordered & 0.1996(29)\\
CLE  & $12^4$ & 5.6 & 0.0 & 0.01 & 0.000294 & 100 & ordered & 0.1985(17)\\
CLE  & $16^4$ & 5.6 & 0.0 & 0.01 & 0.000183 & 5 & ordered & 0.1968(8) \\
CLE  & $16^4$ & 5.6 & 0.0 & 0.005 & 0.000094 & 5 & $dt=0.01$ & 0.1997(9) \\
\hline
\end{tabular}
\caption{Chiral condensate from simulations at $\beta=5.6$, $\mu=0$ for various
types of simulation and choice of parameters.}
\label{tab:pbp}
\end{table}

\begin{table}[htb]
\setlength{\tabcolsep}{10pt}
\begin{tabular}{lllllllll}
\hline
& lattice & $\beta$ & $\mu$ & $dt$ & $dt_{adaptive}$ & cools & start &  
$F(U)$ \\                                                          
\hline                 
CLE  & $12^4$ & 5.6 & 0.0 & 0.01 & 0.000309 & 5 & ordered & 0.1436(28) \\
CLE  & $12^4$ & 5.6 & 0.0 & 0.01 & 0.000299 & 15 & disordered & 0.1479(41) \\
CLE  & $12^4$ & 5.6 & 0.0 & 0.01 & 0.000294 & 100 & ordered & 0.1496(31) \\
CLE  & $16^4$ & 5.6 & 0.0 & 0.01 & 0.000183 & 5 & ordered & 0.1533(15) \\
CLE  & $16^4$ & 5.6 & 0.0 & 0.005 & 0.000094 & 5 & $dt=0.01$ & 0.1487(20) \\
\hline
\end{tabular}
\caption{Unitarity norms $F(U)$ from CLE simulations at $\beta=5.6$, $\mu=0$ 
for various choices of parameters.}
\label{tab:unorm}
\end{table}

There is good agreement between the RLE and the exact RHMC observables when we
used an input $dt=0.01$. We note that changing the number of gauge-cooling
steps has little effect on the chiral condensates, plaquettes and unitarity
norms for the CLE simulations. Neither does changing the starting conditions
for the runs. This was also found to be true at $\mu=0.5$ where we ran CLE
simulations from an ordered start with 5 gauge-cooling steps per update and
from a disordered start with 10 gauge-cooling steps per update. The chiral
condensate shows a small but significant finite volume effect in going from a
$12^4$ to a $16^4$ lattice. Reducing $dt$ from $0.01$ to $0.005$ on the $16^4$
lattice increases the CLE measurement of the chiral condensate. Considering
only the leading dependence on $dt$ and hence on $dt_{adaptive}$ which is
linear, we predict $\langle\bar{\psi}\psi\rangle \approx 0.2025$ in the limit
$dt \rightarrow 0$. We note that the difference between this and the true
(RHMC) value is $\approx 0.013$, compared with the difference between the
$dt=0.01$ value and the true value on the $12^4$ lattice is $\approx 0.014$.
Since we know that one needs a smaller $dt$ on a larger lattice we believe
that $dt=0.01$ is adequate for the $12^4$ lattice, at least at $\mu=0$. For
$\beta=5.7$ on a $16^4$ lattice at $\mu=0$, the RHMC simulations predict that
the chiral condensate, the observable most sensitive to simulation methods and
parameters, $\langle\bar{\psi}\psi\rangle=0.1752(2)$, the RLE simulations with
$dt=0.01$ give $\langle\bar{\psi}\psi\rangle=0.1756(3)$, while the CLE
simulations with $dt=0.01$ and 5 gauge-cools/update yield
$\langle\bar{\psi}\psi\rangle=0.1731(10)$. While the CLE probably still has
some systematic error, we consider it to be small enough to be acceptable.
Hence we choose $dt=0.01$ and 5 gauge-cools/update for all CLE simulations 
except those mentioned in the tables. Here we implicitly assume that adaptive
rescaling of $dt$ is sufficient to tame any wild fluctuations which are
produced by going to non-zero $\mu$.

For comparison, we perform RHMC simulations of the phase-quenched theory with
the same parameters and lattice sizes. At each $\beta$ and $\mu$ we run 10,000
length-1 trajectories (except at $\mu=0$, where we ran 20,000 trajectories). 
The observables for the phase-quenched theory should remain constant at their
$\mu=0$ values up to $\mu \approx m_\pi/2$ in the limit that the explicit
symmetry-breaking term's coefficient $\lambda$ vanishes, up to finite
temperature and volume corrections, whereas in the full theory, these
observables should remain constant until $\mu \approx m_N/3$. (See
\cite{Kogut:2002zg} for definition of $\lambda$). Note that QCD with an isospin
chemical potential $\mu_I$ is identical to the phase-quenched theory with
chemical potential $\mu=\mu_I/2$. This is because the 2-flavour phase-quenched
theory is a theory with 1 quark which is a colour triplet and 1
conjugate-quark which is a colour anti-triplet and has the opposite parity
from the regular quark. These can form a pion-like state of 1 quark and 1
conjugate quark, which can form a colourless quark-number breaking condensate.
If we write $\Psi$ as a `flavour' doublet with components the normal quark and
the anti-conjugate-quark, then the symmetry breaking term in the Lagrangian is
$i\lambda\bar\Psi\gamma_5\tau_2\Psi$.) For $\beta=5.6$, $m=0.025$, $m_\pi/2
\approx 0.21$ and $m_N/3 \approx 0.33$ \cite{Bitar:1990cb} while for
$\beta=5.7$, $m=0.025$, $m_\pi/2 \approx 0.194$ and $m_N/3 \approx 0.28$
\cite{Brown:1991qw,Schaffer:1992rq}, so these 2 transitions should be
distinguishable. At saturation, the quark-number density should be 3 (one
quark of each colour at each site), the chiral condensate should vanish, and
the fermions should decouple from the gauge fields, so gauge observables such
as the plaquette should have their pure gauge values.

\begin{figure}[htb]
\parbox{2.9in}{
\epsfxsize=2.9in
\epsffile{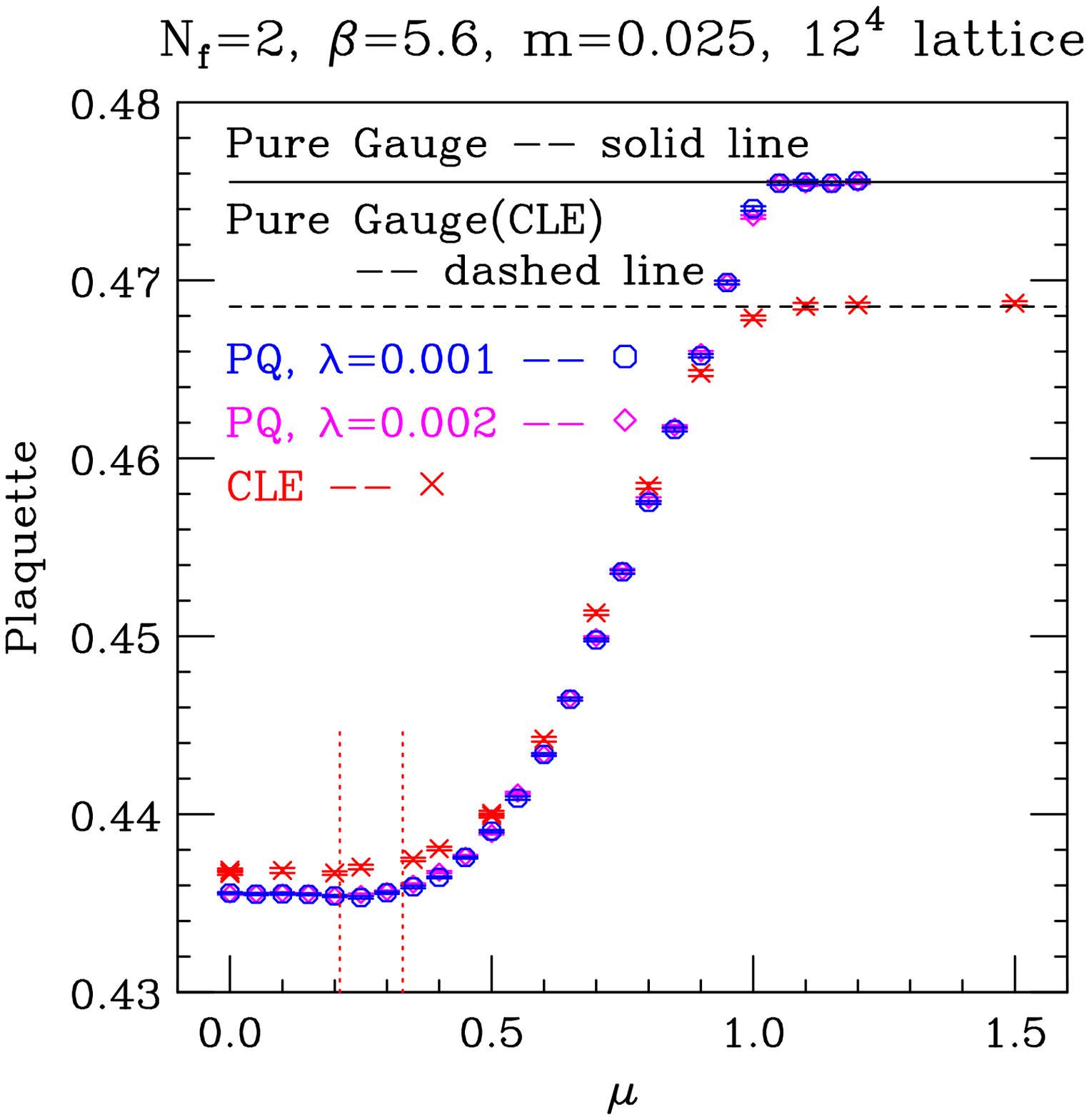}
}
\parbox{0.2in}{}
\parbox{2.9in}{
\epsfxsize=2.9in
\epsffile{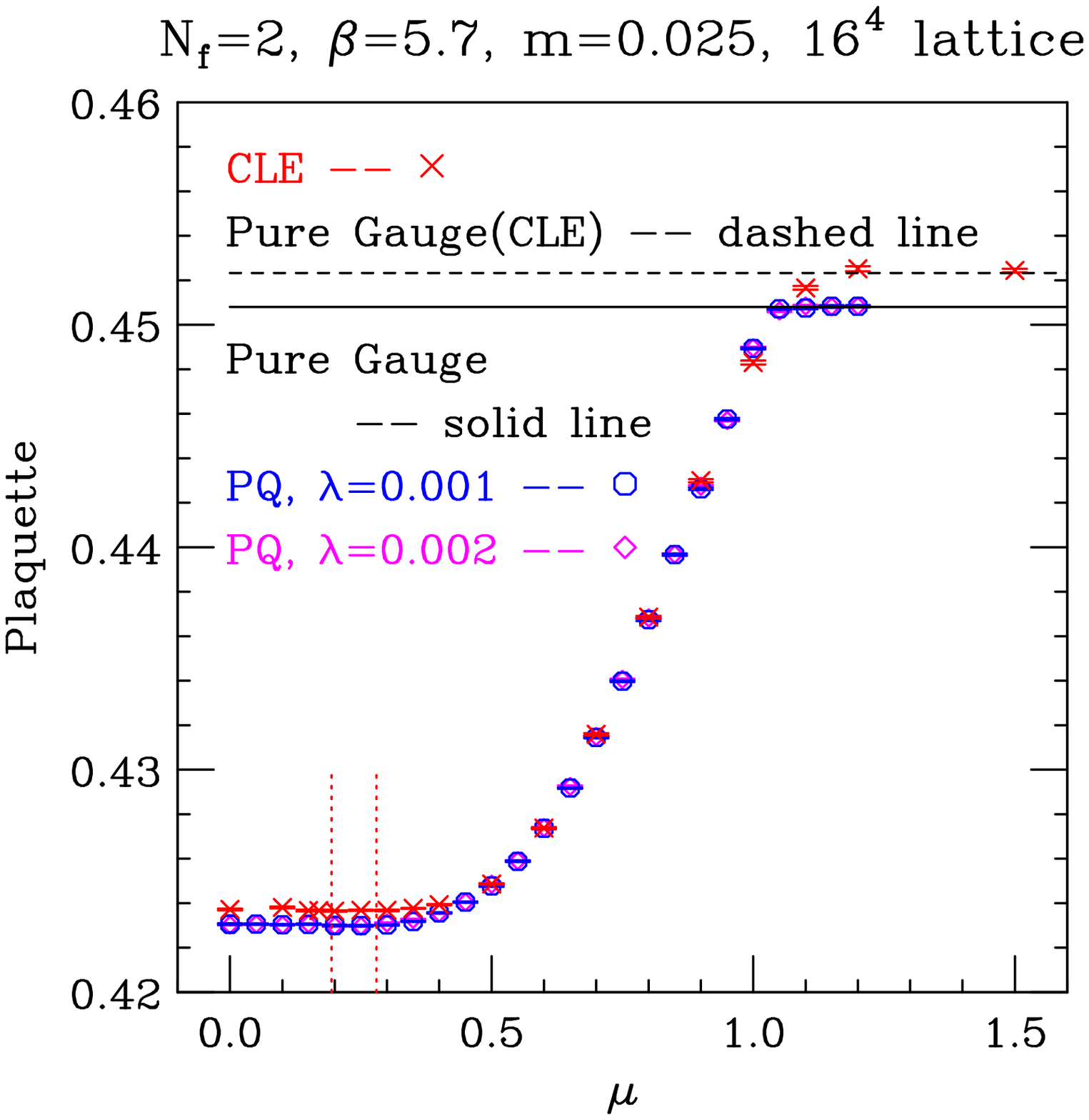}
}
\caption{Plaquettes as functions of $\mu$ for (a) $\beta=5.6$, $m=0.025$ on a
$12^4$ lattice and for (b) $\beta=5.7$, $m=0.025$ on a $16^4$ lattice. Both
complex Langevin (CLE) and phase-quenched (PQ) results are presented. Vertical
dotted lines are at $\mu=m_\pi/2$ and $\mu=m_N/3$.}
\label{fig:plaquette}
\end{figure}

Figure~\ref{fig:plaquette} shows the average plaquettes as functions of $\mu$
for $\beta=5.6$, $m=0.025$ on a $12^4$ lattice and $\beta=5.7$, $m=0.025$ on
a $16^4$ lattice, from our CLE simulations. The corresponding results for
the phase-quenched theory are shown for 2 different values of the symmetry 
breaking parameter $\lambda$ indicating that there is relatively little
$\lambda$ dependence. For $\mu=0$ at $\beta=5.6$, while the difference between
the CLE value and the exact value is not large, it is significant. At 
$\beta=5.7$ the relative difference is almost a factor of 2 smaller. At
saturation, for both $\beta$s, the plaquette for the phase-quenched theory is
identical to that of the pure gauge theory within statistical errors. For both
$\beta$s the plaquette value for the full theory at saturation predicted by
the CLE, agrees within statistical errors with that predicted by CLE
simulations of the pure gauge theory. The values predicted by the CLE for
pure gauge theory differ significantly from the correct value (see section~4).
However, this difference is much smaller for $\beta=5.7$ than for $\beta=5.6$.

\begin{figure}[htb]  
\parbox{2.9in}{      
\epsfxsize=2.9in     
\epsffile{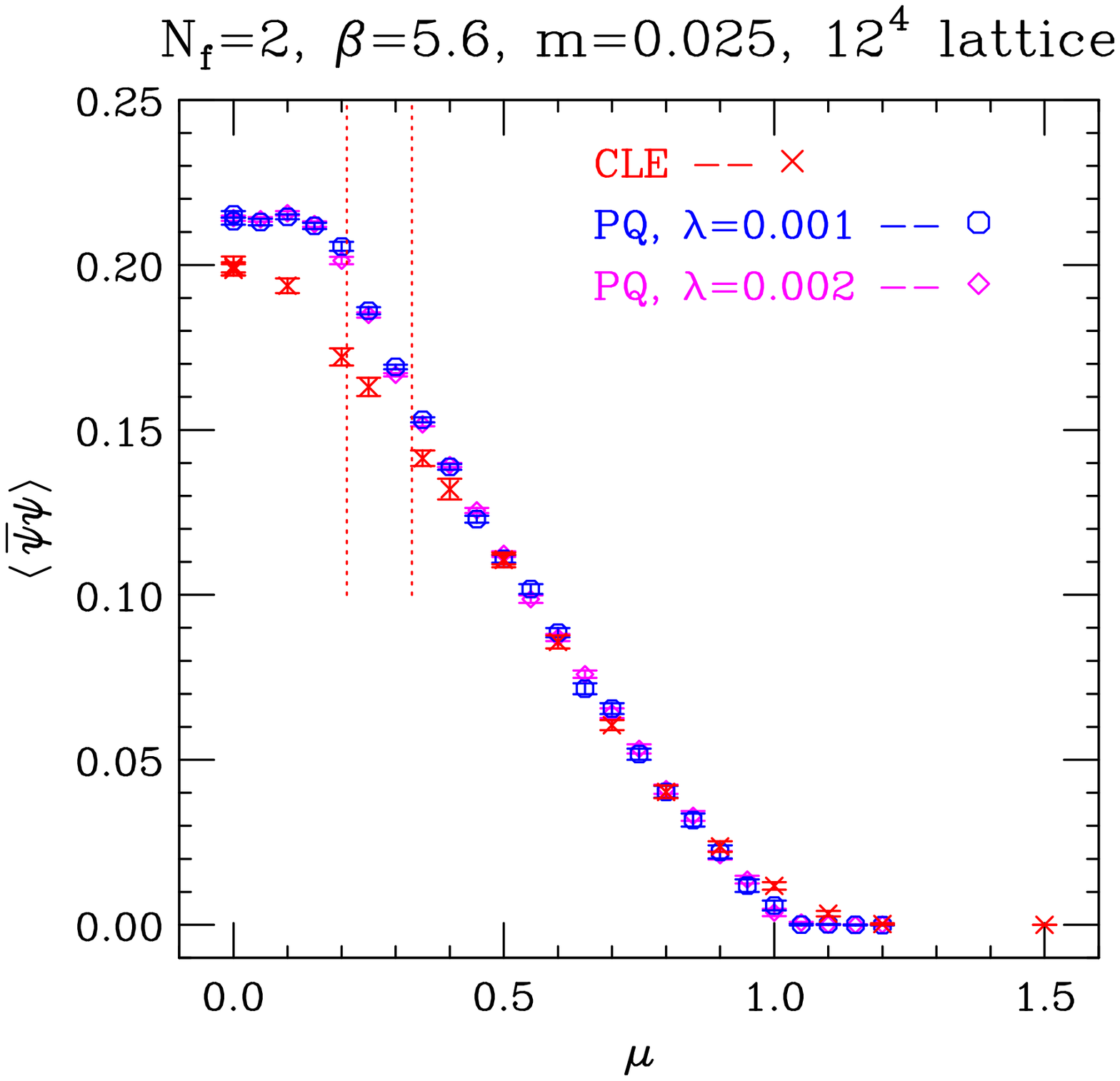}
}                  
\parbox{0.2in}{}               
\parbox{2.9in}{                   
\epsfxsize=2.9in               
\epsffile{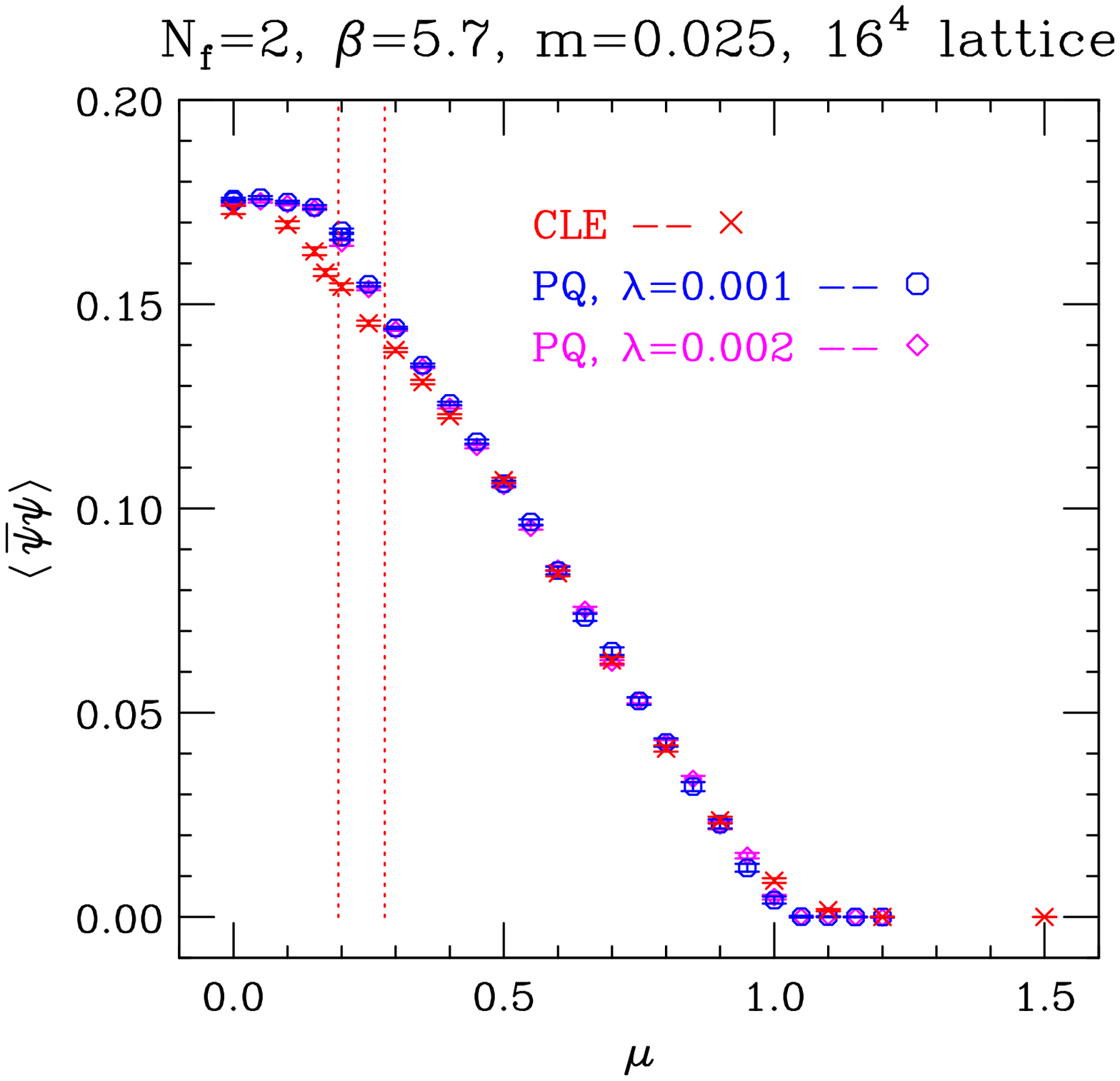}
}                             
\caption{Chiral condensates as functions of $\mu$ for (a) $\beta=5.6$,
$m=0.025$ on a $12^4$ lattice and for (b) $\beta=5.7$, $m=0.025$ on a $16^4$
lattice. Both complex Langevin (CLE) and phase-quenched (PQ) results are
presented. Vertical dotted lines are at $\mu=m_\pi/2$ and $\mu=m_N/3$.}
\label{fig:pbp}                                       
\end{figure} 

The chiral condensates for $\beta=5.6$ and $\beta=5.7$ are shown as functions
of $\mu$ in figure~\ref{fig:pbp} for both our CLE simulations of QCD at
finite $\mu$ and RHMC simulations of the phase-quenched theory. At $\mu=0$
the CLE value of this condensate at $\beta=5.6$ is $\approx$~6.6\% too low, 
while at $\beta=5.7$ it is $\approx$~1.2\% too low, a considerable improvement.
As $\mu$ is increased from zero, the condensates for both $\beta$s start to
fall for $\mu < m_\pi/2$ rather than for $\mu \approx m_N/3$. Therefore, even
though $\beta=5.7$ shows an improvement over $\beta=5.6$, it is still a worse
approximation to the physics than is the phase-quenched theory. It is still
unclear if going to a much weaker coupling will produce the correct results or
those of the phase-quenched theory. At both $\beta$s the condensate reaches
its saturation value of zero at large $\mu$.

\begin{figure}[htb]  
\parbox{2.9in}{      
\epsfxsize=2.9in     
\epsffile{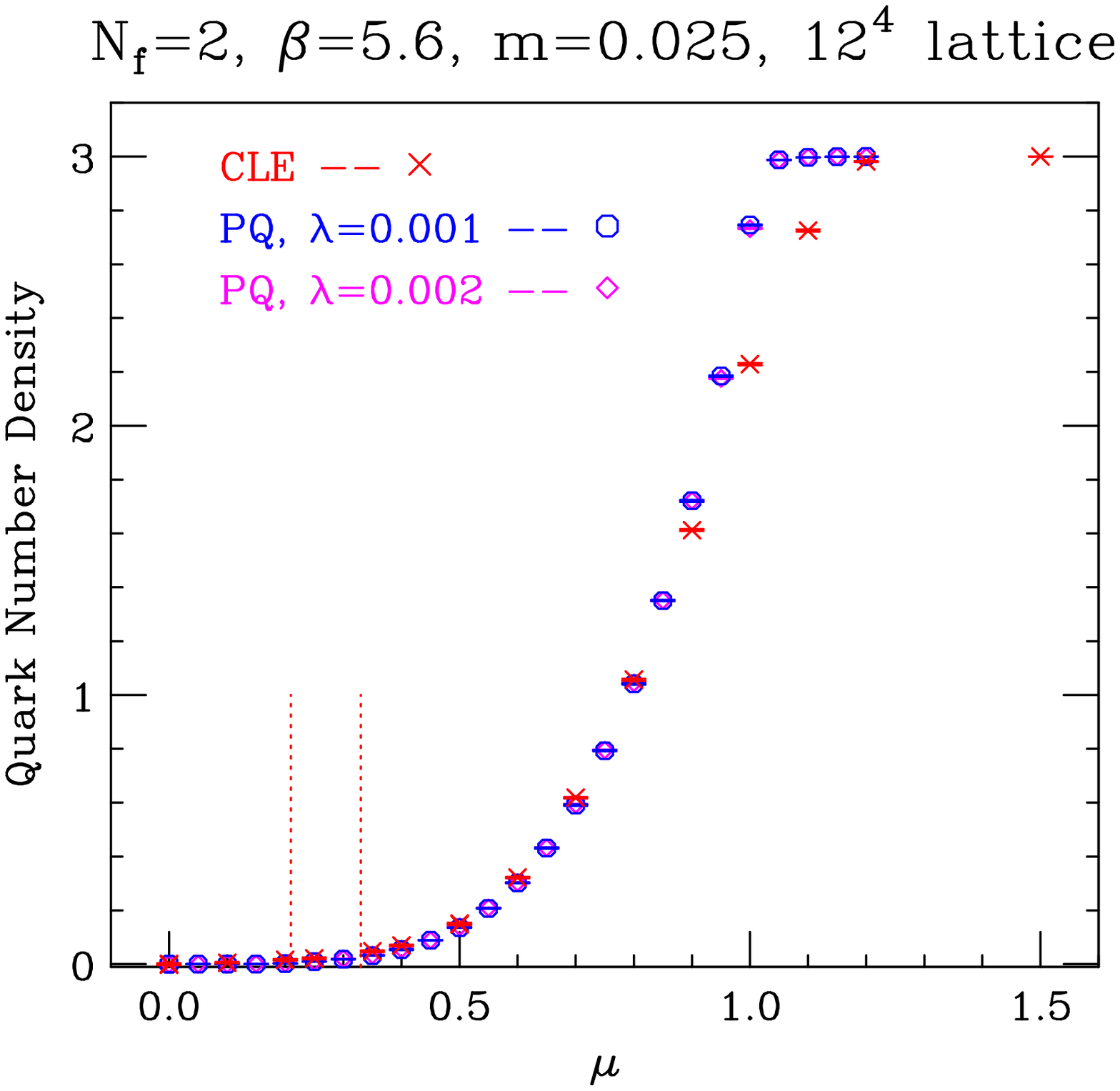}
}                  
\parbox{0.2in}{}               
\parbox{2.9in}{                   
\epsfxsize=2.9in               
\epsffile{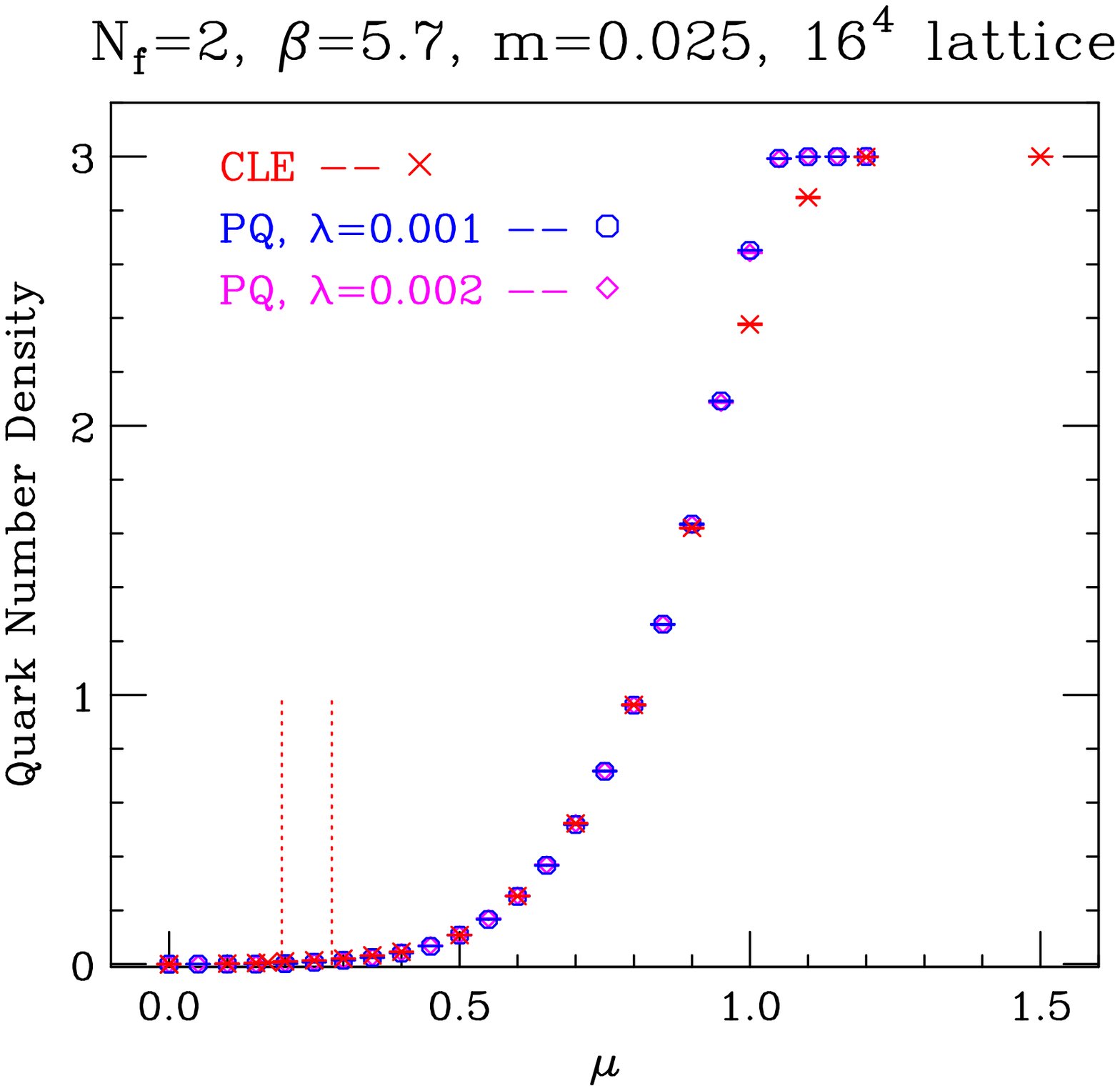}      
}                             
\caption{Quark-number densities as functions of $\mu$ for (a) $\beta=5.6$,
$m=0.025$ on a $12^4$ lattice and for (b) $\beta=5.7$, $m=0.025$ on a $16^4$
lattice.  Both complex Langevin (CLE) and phase-quenched (PQ) results are
presented. Vertical dotted lines are at $\mu=m_\pi/2$ and $\mu=m_N/3$.}       
\label{fig:qnd}                                             
\end{figure}  

\begin{figure}[htb]
\parbox{2.9in}{
\epsfxsize=2.9in
\epsffile{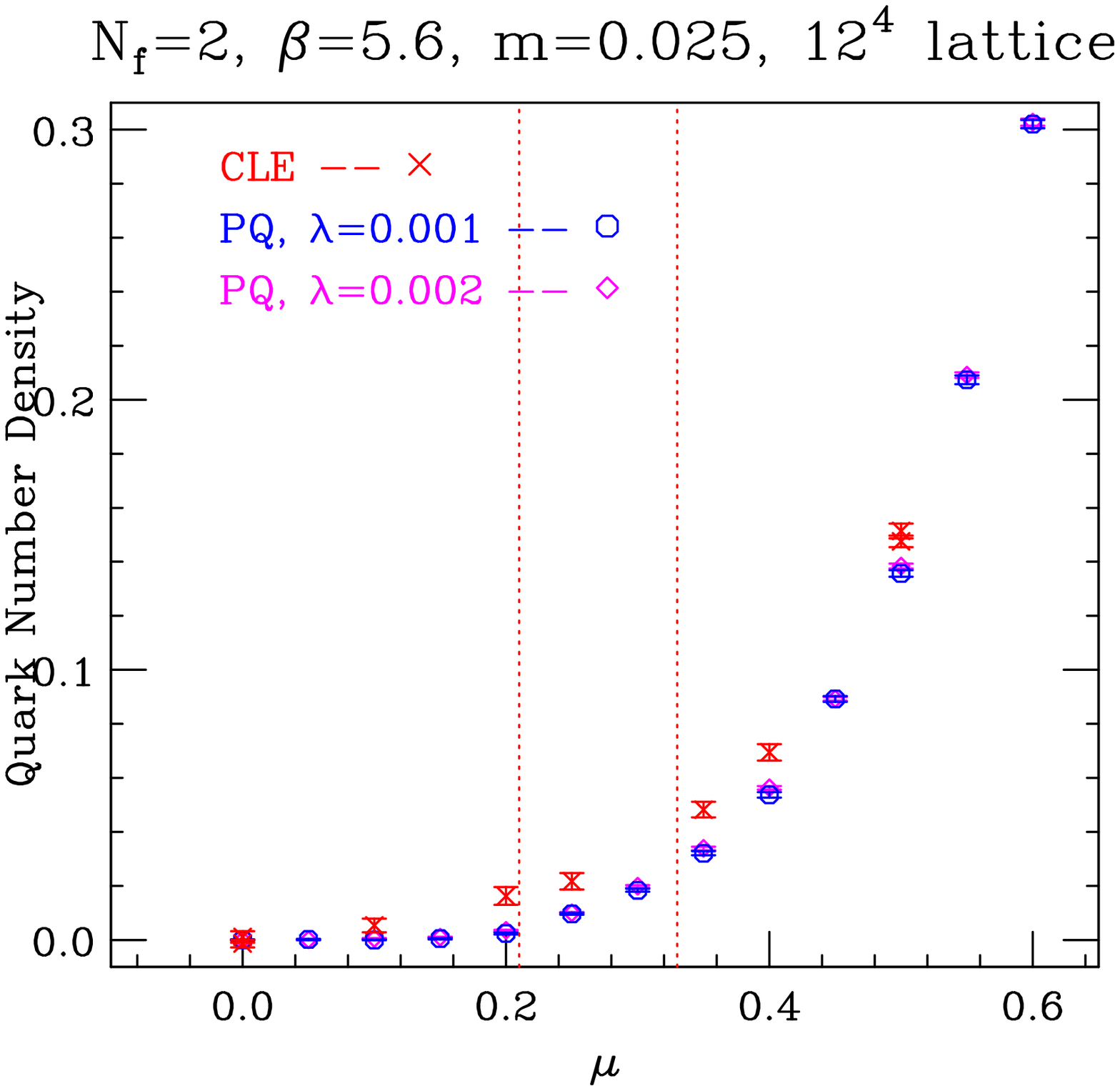}
}
\parbox{0.2in}{}
\parbox{2.9in}{
\epsfxsize=2.9in
\epsffile{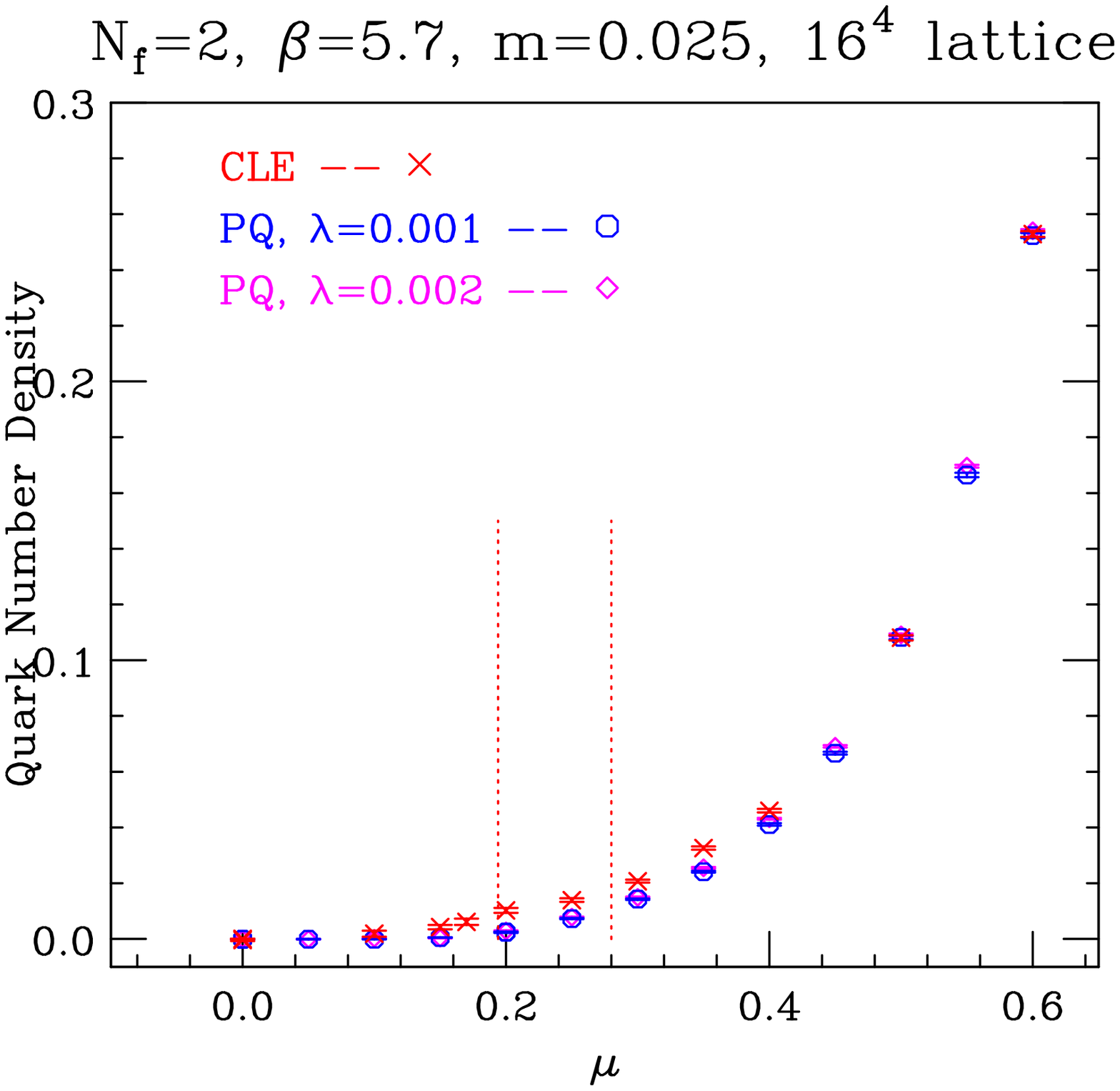}
}
\caption{As for figure~\ref{fig:qnd}, but showing the low $\mu$ region on an
expanded scale.}
\label{fig:qnd_zoom}                                                     
\end{figure}

Figure~\ref{fig:qnd} shows the quark-number densities for our $\beta=5.6$ and
$\beta=5.7$ simulations. On this scale, the CLE results for the full theory
and RHMC results for the phase-quenched theory look almost identical, except
on their approach to saturation. Figure~\ref{fig:qnd_zoom}, shows an expanded
version of the low $\mu$ region which indicates that the apparent agreement is
only because the number densities are so close to zero for small $\mu$. The
results for the 2 theories are closer for $\beta=5.7$ than for $\beta=5.6$.
This would be expected if the CLE produces phase-quenched results in the
weak-coupling limit, but also could indicate that it produces correct results
in that limit. Note that it is difficult, if not impossible, to determine the
positions of the transitions from these quark-number density graphs.

\begin{figure}[htb]
\epsfxsize=4in
\centerline{\epsffile{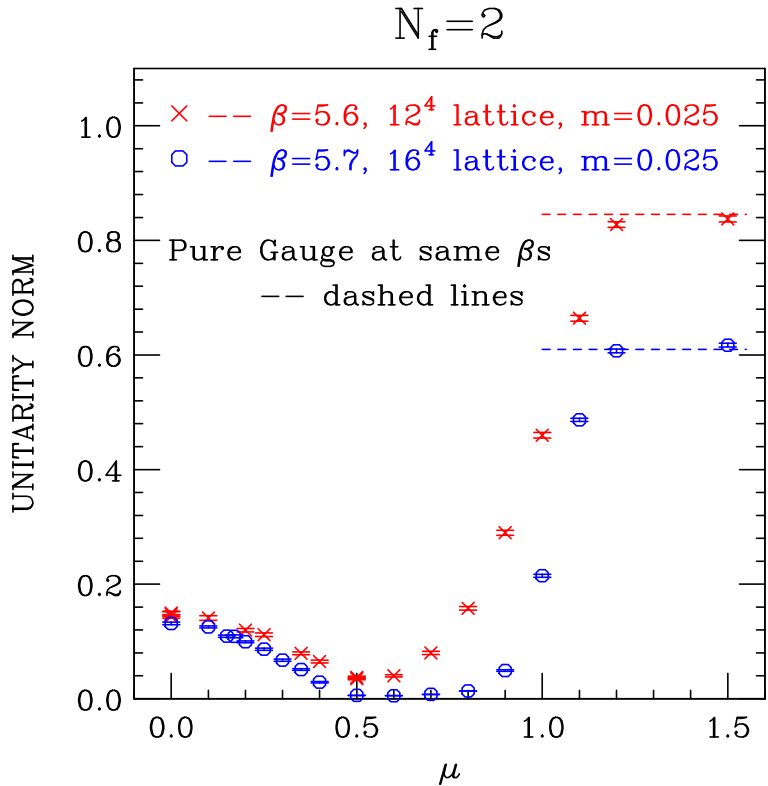}}
\caption{Average unitarity norms as functions of $\mu$ for $\beta=5.6$,
$m=0.025$ and $\beta=5.7$, $m=0.025$.}
\label{fig:unorm}
\end{figure}

Figure~\ref{fig:unorm} shows the average unitarity norms for the CLE
simulations at $\beta=5.6$ and $\beta=5.7$ as functions of $\mu$. In both 
cases this norm decreases from its value at $\mu=0$ as $\mu$ is increased 
reaching a minimum around $0.5$ or $0.6$ before increasing to a maximum at
saturation, which is the value for CLE simulations of the pure gauge theory
at the same $\beta$ value. The values of the unitarity norms for $\beta=5.7$
lie below the values of those at $\beta=5.6$ for the same $\mu$. It has been
suggested, on the basis of simulations of lattice QCD in the heavy-dense limit
that there is some value of the unitarity norm around $0.1$ below which the
CLE will produce correct results \cite{Aarts:2016qrv}
\footnote{We have investigated the suggestion from this paper that one should
terminate the simulation at a given $\beta$ and $\mu$ before the fluctuations
in the observables increase substantially. For our simulations, while this
improves the results at $\mu=0$, it does not prevent the precocious onset of
the transition. In addition, such restricted simulations are short and their
length decreases with increasing $\mu$, essentially vanishing above the
transition, so it is doubtful that the system has time to equilibrate.}. 
From our simulations at $m=0.025$ there does appear to be such a value for
$\mu=0$, between $0.13$ and $0.15$. However, if there is such a value for $\mu
> 0$, it decreases with increasing $\mu$, lying below the value of this norm
for $\beta=5.7$, at least through the transition region.

At $\beta=5.7$, the value of the unitarity norm drops by roughly a factor of
2 from $\mu=0$ where it is $\approx 0.131$ to $\mu=0.3$ at the upper end of 
the transition region, where it is  $\approx 0.067$. From there it falls by an
order of magnitude, reaching a minimum between $\mu=0.5$ where it is 
$\approx 0.0058$ and $\mu=0.6$ where it is $\approx 0.0054$. It is tempting to
suggest that the CLE will produce correct results for $\mu$s around this
rather broad minimum. Moreover, for $0.5 \le \mu \le 0.9$ the plaquette, the
chiral condensate, and the quark-number density are in good agreement with
those of the phase-quenched approximation. (Above $\mu=0.9$ we are in the
regime controlled by saturation a lattice artifact.) Either, in the large $\mu$
region, the full and phase-quenched theories give the same physics or this is
a sign that the CLE breakdown produces phase-quenched results as suggested by
random matrix theory. We shall have more to say about this shortly.

\begin{figure}[htb]
\parbox{2.9in}{
\epsfxsize=2.9in
\epsffile{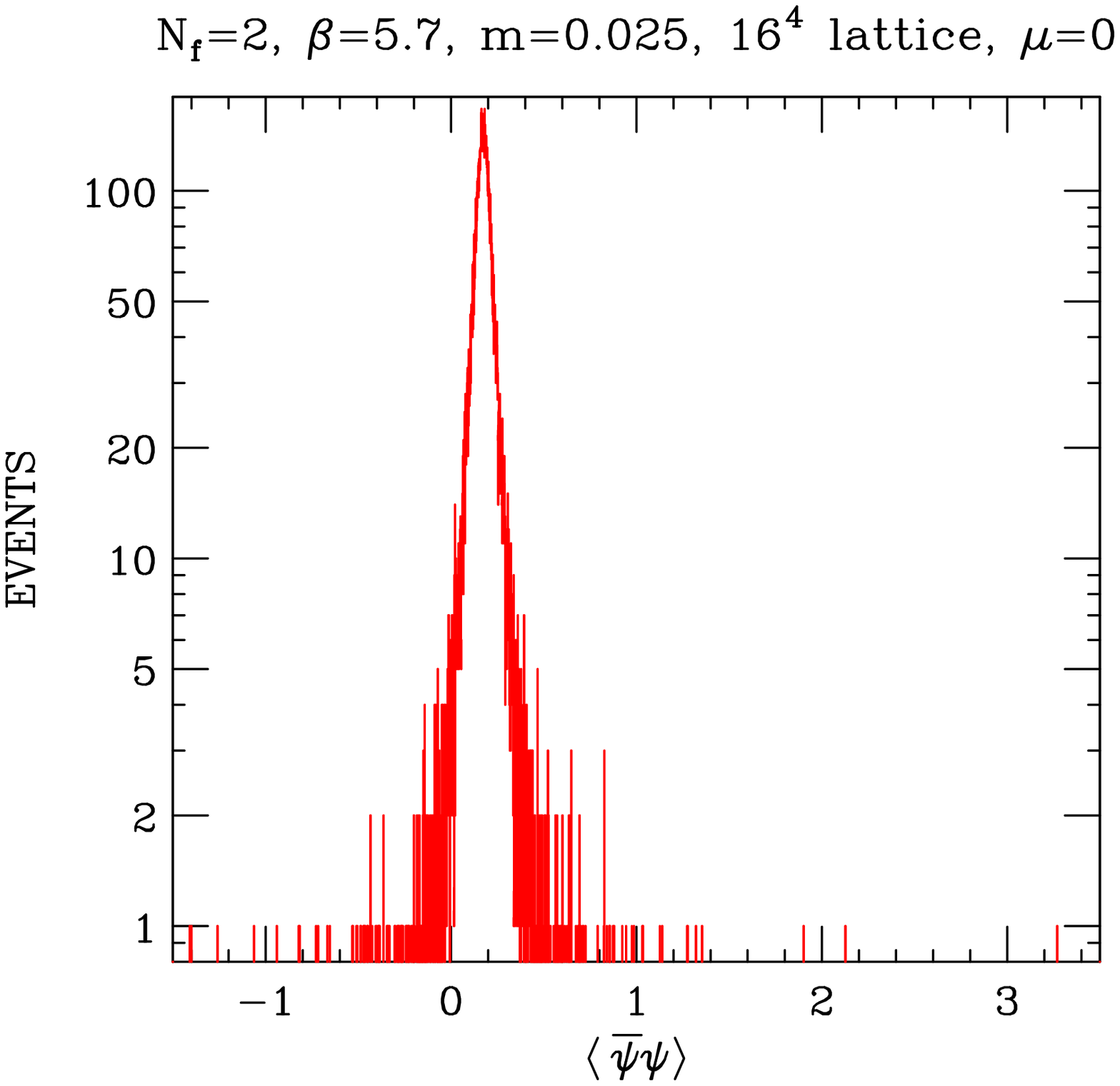}
}
\parbox{0.2in}{}
\parbox{2.9in}{
\epsfxsize=2.9in
\epsffile{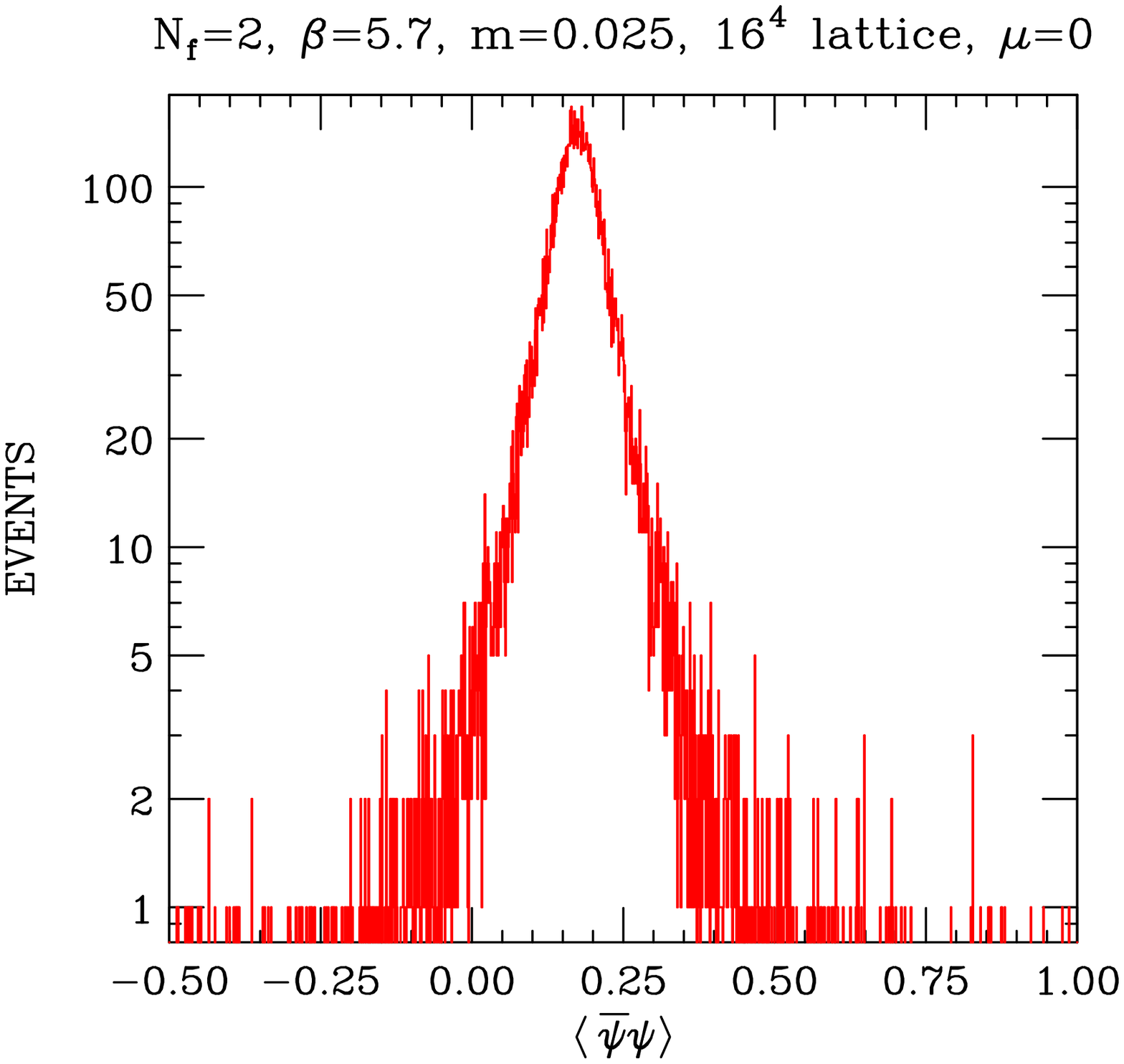}
}
\caption{Histogram of values of the chiral condensate from a CLE simulation   
at $\beta=5.7$, $m=0.025$, $\mu=0$. a) Full histogram. b) Central portion of
this histogram.}
\label{fig:pbp0}
\end{figure}

\begin{figure}[htb]                                                             
\parbox{2.9in}{    
\epsfxsize=2.9in
\epsffile{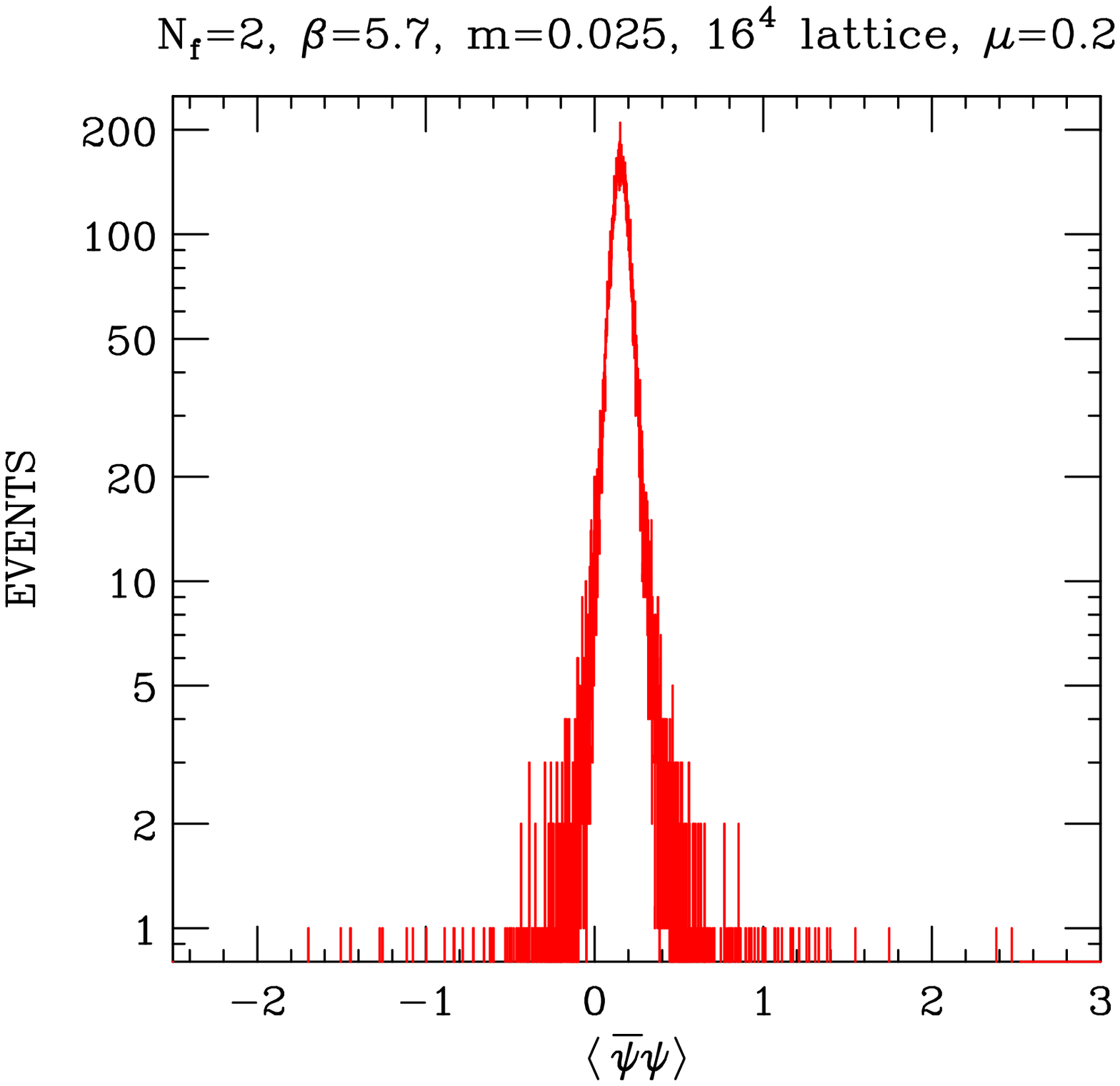}
}                                    
\parbox{0.2in}{}
\parbox{2.9in}{ 
\epsfxsize=2.9in
\epsffile{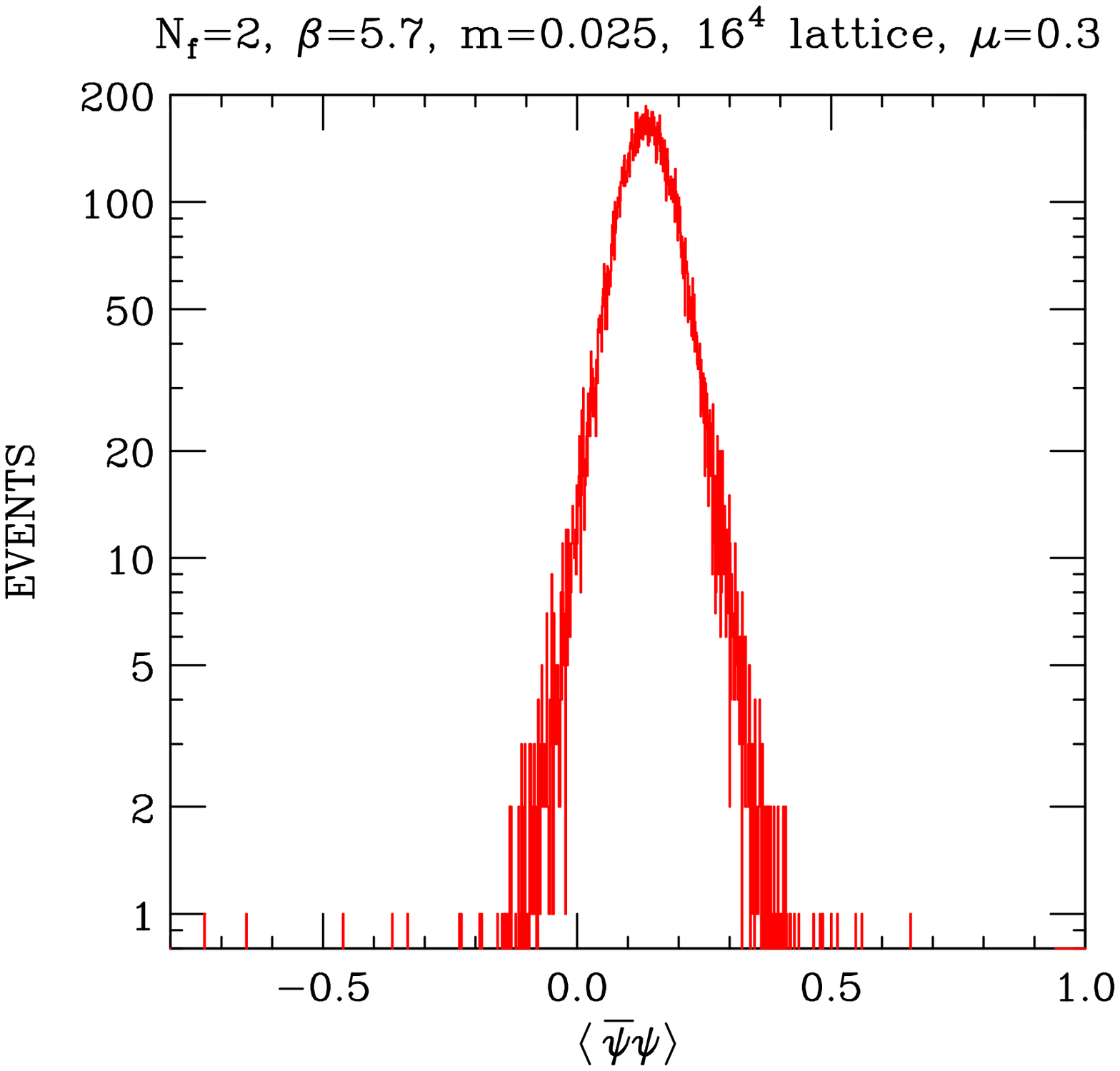}
}                                         
\caption{Histogram of values of the chiral condensate from a CLE simulation
at $\beta=5.7$, $m=0.025$. a) $\mu=0.2$. b) $\mu=0.3$.}
\label{fig:pbp23}
\end{figure} 

\begin{figure}[htb]                          
\parbox{2.9in}{                                                                 
\epsfxsize=2.9in   
\epsffile{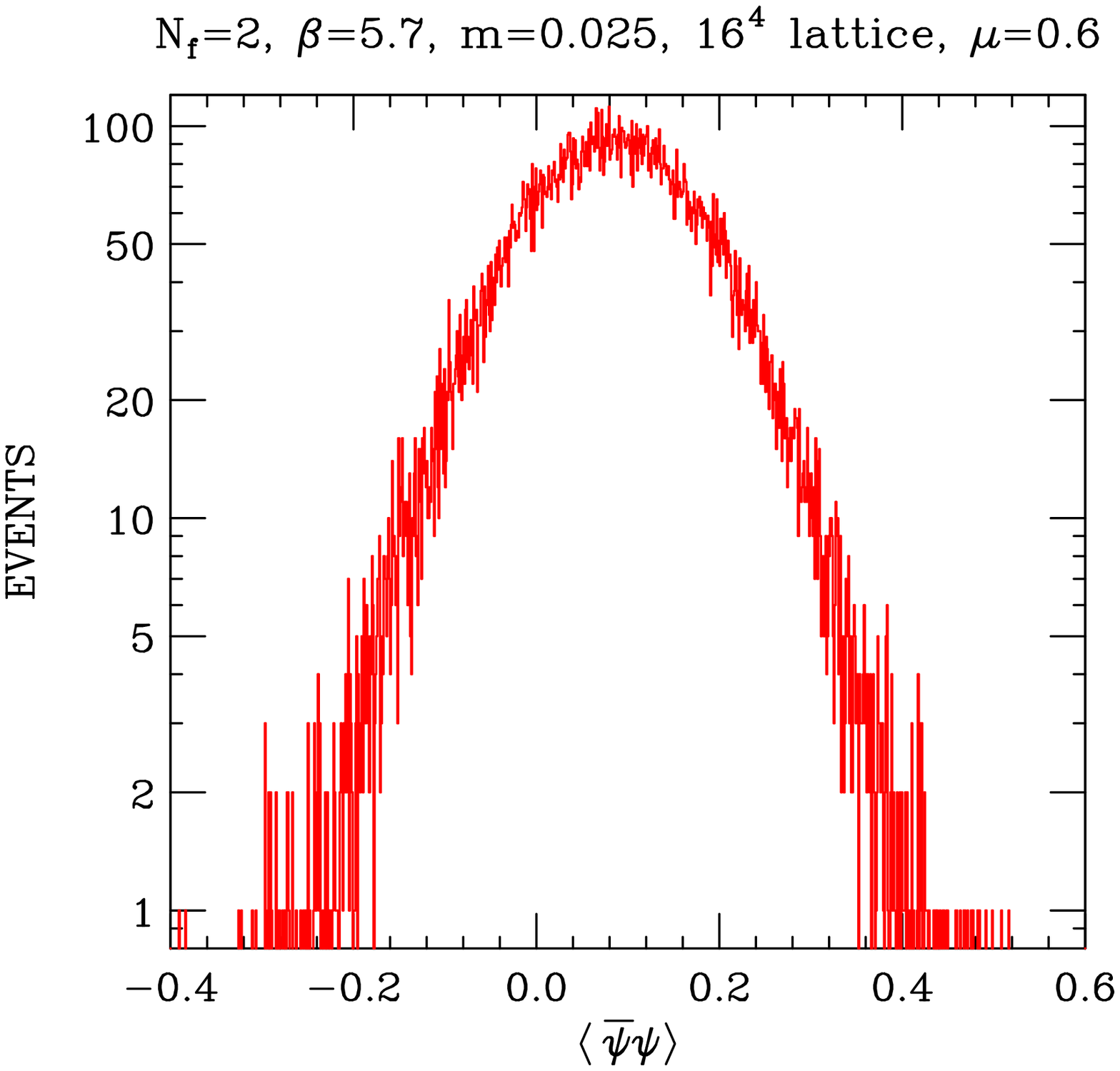}
}                                    
\parbox{0.2in}{}                     
\parbox{2.9in}{ 
\epsfxsize=2.9in
\epsffile{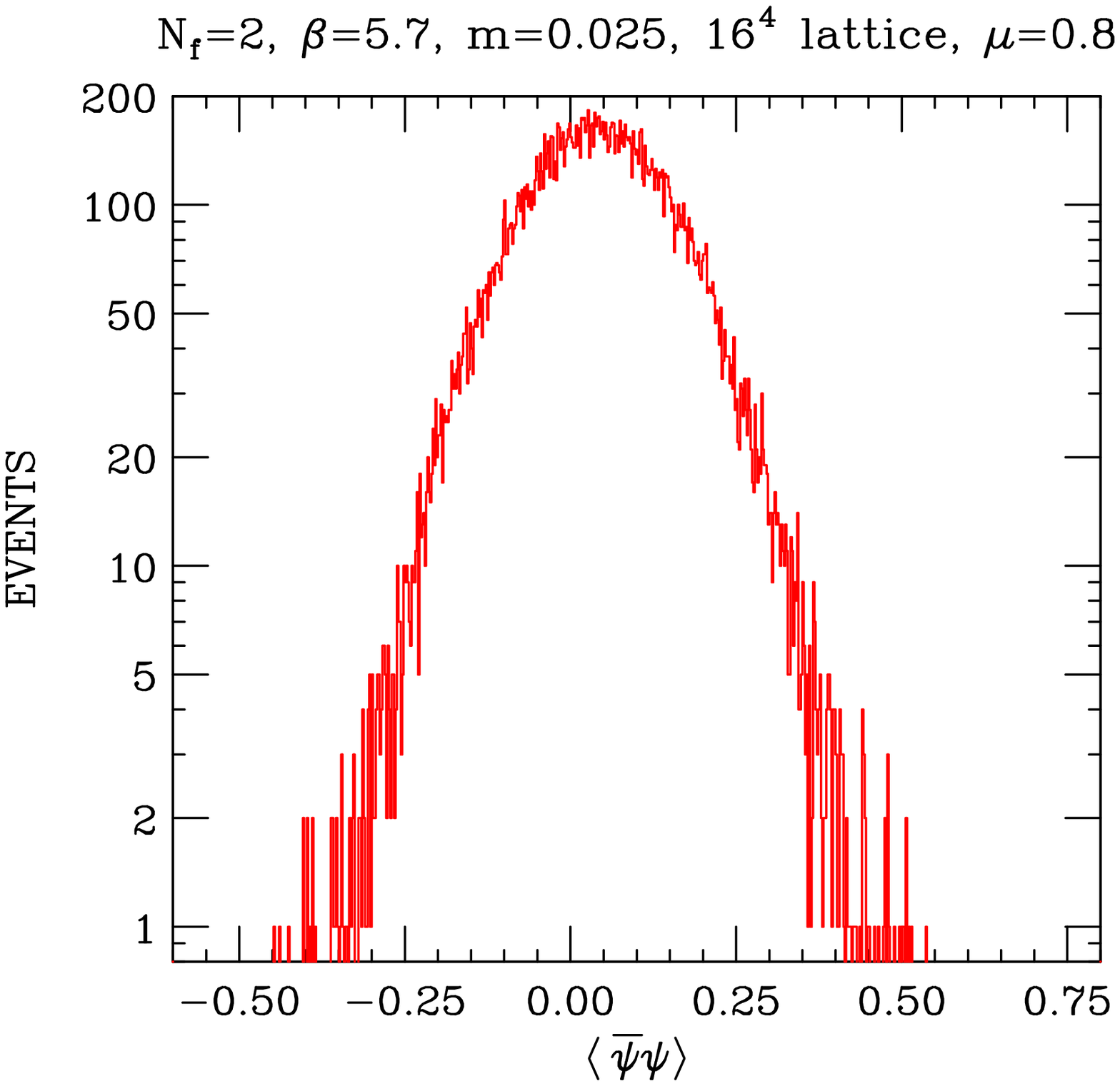}
}                                    
\caption{Histogram of values of the chiral condensate from a CLE simulation
at $\beta=5.7$, $m=0.025$. a) $\mu=0.6$. b) $\mu=0.8$.}                        
\label{fig:pbp68}                                                               
\end{figure}

To test whether the assumption that the CLE is valid for $\beta=5.7$, $m=0.025$
for $\mu$ near the unitarity-norm minimum is reasonable, even though it fails
for small $\mu$, we examine the distribution of values of the chiral
condensate for $0 \le \mu \le 0.9$. (For $\mu > 0.9$ the system is influenced
by saturation, a lattice artifact.) The chiral condensate is chosen since it
has poles at the same places as those of the drift term. Hence if these poles
are approached too closely, which invalidates the CLE, it should show large
(non-Gaussian) excursions in its distribution. Figure~\ref{fig:pbp0} presents
the histogram of values of $\langle\bar{\psi}\psi\rangle$ at $\mu=0$, which is
typical of the distributions for small $\mu$. This histogram has long tails,
with a few outliers, which indicates that the poles in the drift term as well
as those in $\bar{\psi}\psi$ are being approached. The fact that good results
are obtained for $\mu=0$ indicates that although the trajectory of the system
approaches the poles this does not necessarily cause a breakdown of the CLE.
However, as $\mu$ is increased, and the tails slowly become more prominent,
the CLE does start to deviate from the correct physics. In
figure~\ref{fig:pbp23} we show histograms at $\mu=0.2$, close to $m_\pi/2$ and
hence at the beginning of the transition region if the physics were that of
the phase-quenched theory, and at $\mu=0.3$ close to $m_N/3$ and thus to the
transition expected for QCD at finite $\mu$. At $\mu=0.2$ the tails appear to
be close to maximal. Beyond this, they decrease, and are noticeably smaller at
$\mu=0.3$. Figure~\ref{fig:pbp68} shows the histograms for $\mu=0.6$ at the
minimum of the unitarity norm, and $\mu=0.8$ in the large $\mu$ regime. By
$\mu=0.6$ the non-gaussian tails have almost vanished and remain insignificant
over the range $0.5 \le \mu \le 0.9$ as can be seen in the histogram at
$\mu=0.8$. In this high $\mu$ regime we have also looked at the histograms of
quark-number density distributions (since the quark-number operator also has
poles at the zeros of the fermion determinant), which are also well-behaved
over this range of $\mu$s.

\begin{figure}[htb]
\parbox{2.9in}{
\epsfxsize=2.9in
\epsffile{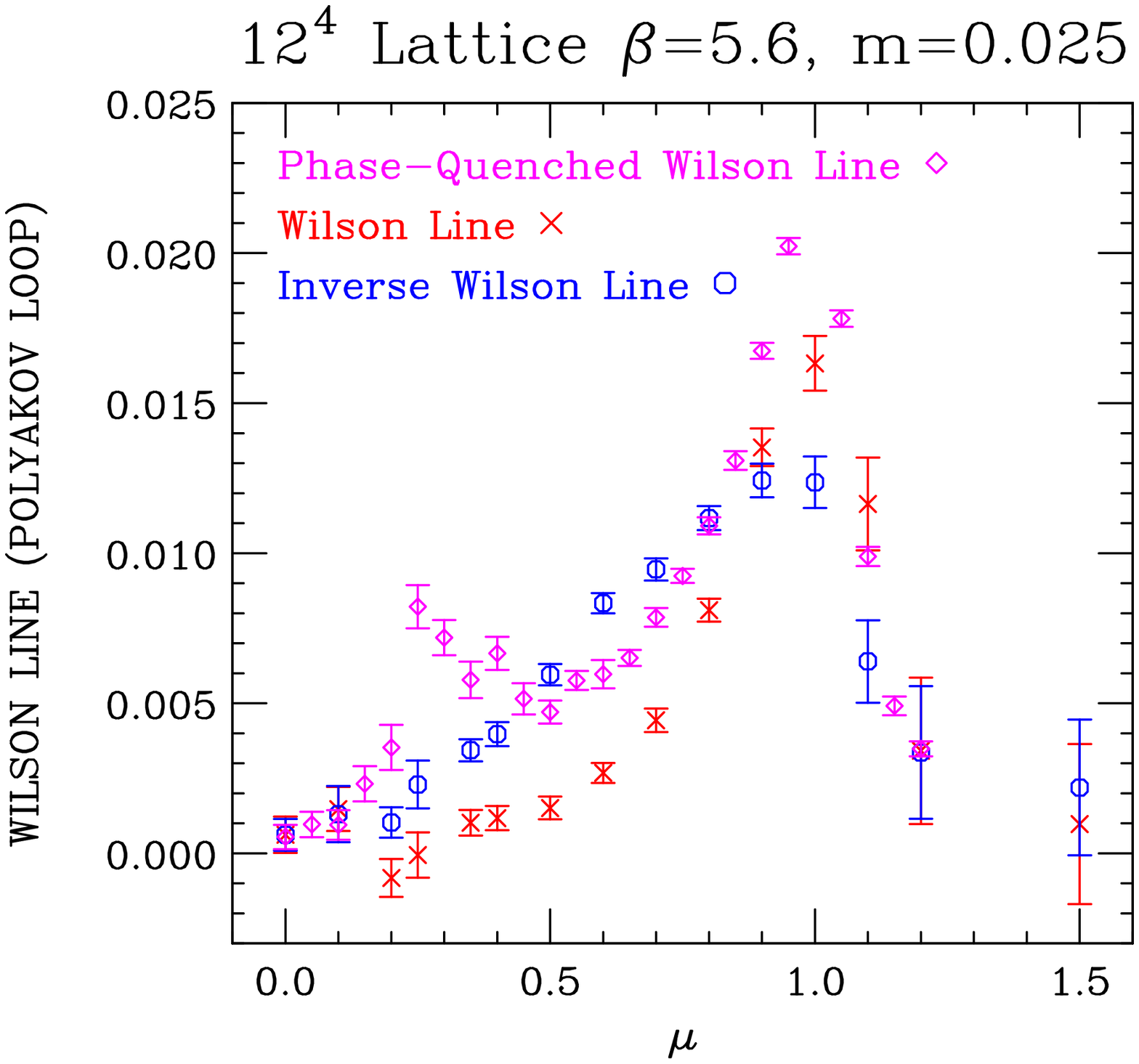}
}
\parbox{0.2in}{}
\parbox{2.9in}{
\epsfxsize=2.9in
\epsffile{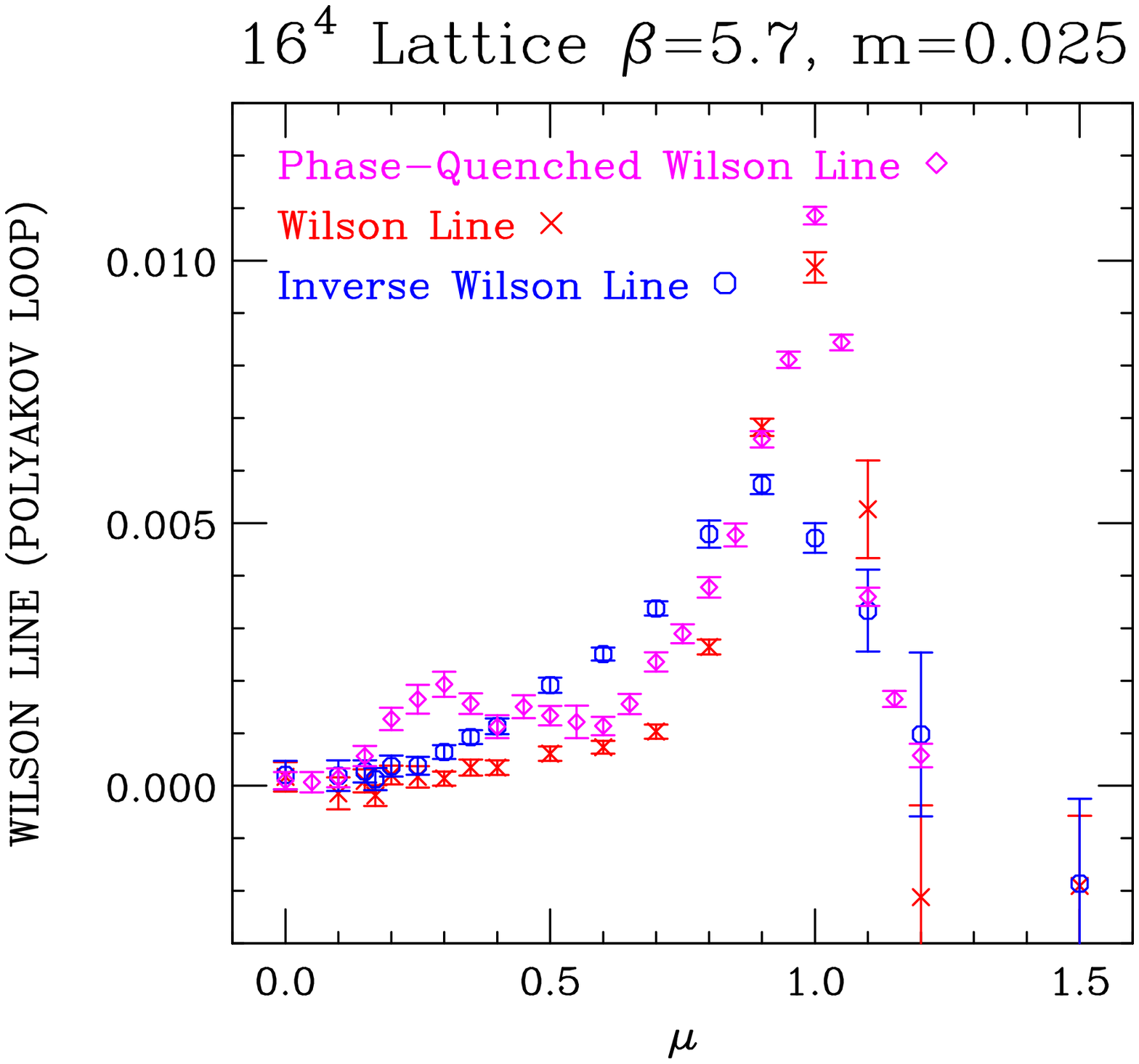}
}
\caption{Wilson Line (Polyakov Loop) and Inverse Wilson Line as functions of 
$\mu$ (a) for $\beta=5.6$, $m=0.025$ on a $12^4$ Lattice and (b) for 
$\beta=5.7$, $m=0.025$ on a $16^4$ Lattice} 
\label{fig:wilson}
\end{figure}

Figure~\ref{fig:wilson} shows the real parts of the Wilson Lines (Polyakov 
Loops) and the Inverse Wilson Lines as functions of $\mu$ for our $\beta=5.6$,
and $\beta=5.7$ from our CLE simulations. The imaginary parts of these
quantities are small, consistent with zero. We include the Wilson Lines from
the corresponding phase-quenched RHMC simulations with $\lambda=0.001$ for
comparison. What we notice is that once $\mu$ is above the transition, if not
before, the Wilson lines start to diverge from their inverses. This means that
the fermion determinant is complex. The fact that the inverse Wilson line lies
above the Wilson line up to some $\mu$ between $0.8$ and $0.9$ indicates that
the phase of the Wilson line and that of the fermion determinant are
positively correlated. Such a correlation was anticipated in
\cite{deForcrand:1999cy}, and observed in simulations of heavy-dense lattice
QCD \cite{Seiler:2012wz,Aarts:2008rr,Langelage:2014vpa,Rindlisbacher:2015pea}. 
It is interesting to note that the crossover point where the phase of the
determinant vanishes lies near to the quark-number density of 1.5 where the
fermi states are half-filled. This behaviour was predicted and observed in
heavy-dense lattice QCD \cite{Rindlisbacher:2015pea}.

The Wilson line and the inverse-Wilson line increase with increasing $\mu$
reaching a maximum just before the effects of saturation start to be felt. At
$\beta=5.7$ the fact that local observables measured in CLE simulations agree
with phase-quenched results, despite the fact that the fermion determinant is
complex, is possible because the phase of the fermion determinant can be
determined by a few low lying eigenvalues of the Dirac operator, whereas other
quantities are determined by the distribution of eigenvalues. This was
discussed most clearly in \cite{Bloch:2017sex}. We also note that the CLE
Wilson and inverse-Wilson lines do not show any significant effect from the
pseudo-transition at $\mu=m_\pi/2$, unlike the Wilson line for the
phase-quenched theory, which might indicate that the CLE will not eventually
yield phase-quenched results in the transition region.

Because the presence of the chemical potential introduces an asymmetry between
space and time which tends to amplify the effects of temperature, the question
arises as to whether the $12^4$ lattice for $\beta=5.6$ and the $16^4$ lattice
for $\beta=5.7$, while being good approximations to zero temperature at $\mu=0$,
remain good approximations to zero temperature as $\mu$ is increased. The
increase in value of the Wilson Line as $\mu$ is increased, discussed in the
previous paragraph, emphasizes the possibility that some of the $\mu$ dependence
of observables other than the Wilson line could actually be finite temperature
effects. To check this we performed test CLE runs on a $16^3 \times 36$ lattice
with $\beta=5.7$ and $m=0.025$ at $\mu=0.2$ and $\mu=0.3$, which bracket the
transition region, and at $\mu=0.8$, a large value of $\mu$, but not so large
as to be influenced by saturation. At all 3 of these $\mu$ values the Wilson
lines are consistent with zero. The values of the the plaquette, chiral
condensate, and quark-number density and unitarity norm measured in these
simulations are compared with those obtained for the corresponding $16^4$
lattice simulations in table~\ref{tab:large-N_t}.

\begin{table}[htb]
\setlength{\tabcolsep}{10pt}
\parbox{2.9in}{
\begin{tabular}{lll}
\hline
\multicolumn{3}{c}{plaquette} \\
\hline
$\mu$ & $16^4$       & $16^3 \times 36$ \\
\hline
0.2   & 0.42365(4)   & 0.42364(3)       \\
0.3   & 0.42368(4)   & 0.42365(3)       \\
0.8   & 0.43686(6)   & 0.43706(3)       \\
\hline
\end{tabular}
}
\parbox{0.2in}{}
\parbox{2.9in}{                   
\begin{tabular}{lll}                                                    
\hline                      
\multicolumn{3}{c}{$\langle\bar{\psi}\psi\rangle$}  \\  
\hline                                                                         
$\mu$ & $16^4$       & $16^3 \times 36$ \\
\hline
0.2   & 0.1543(8)    & 0.1540(3)        \\
0.3   & 0.1388(5)    & 0.1385(3)        \\
0.8   & 0.0412(8)    & 0.0437(9)        \\
\hline
\end{tabular}
}
\parbox{2.9in}{
\begin{tabular}{lll}
\hline
\multicolumn{3}{c}{quark number density} \\
\hline
$\mu$ & $16^4$       & $16^3 \times 36$ \\
\hline
0.2   & 0.0103(9)    & 0.0092(3)        \\
0.3   & 0.0208(6)    & 0.0214(4)        \\
0.8   & 0.9634(16)   & 0.9620(17)       \\
\hline
\end{tabular}
}
\parbox{0.2in}{}
\parbox{2.9in}{                   
\begin{tabular}{lll}                                                    
\hline                      
\multicolumn{3}{c}{unitarity norm} \\
\hline                                                                         
$\mu$ & $16^4$       & $16^3 \times 36$ \\
\hline
0.2   & 0.0994(15)   & 0.0999(12) \\
0.3   & 0.0673(19)   & 0.0667(12) \\
0.8   & 0.0135(3)    & 0.0135(2)  \\
\hline
\end{tabular}
}
\caption{Comparison between local observables from CLE simulations at 
$\beta=5.7$, $m=0.025$ on $16^4$ and $16^3 \times 36$ lattices for 
$\mu=0.2, 0.3, 0.8$.}
\label{tab:large-N_t}
\end{table}
For $\mu=0.2, 0.3$ there is good agreement between these local observables for
$N_t=16$ and $N_t=36$. For $\mu=0.08$ there are small but noticeable
differences in the plaquettes and chiral condensates between the 2 lattice
sizes, while the quark-number densities and unitarity norms are in good
agreement. We conclude that the finite temperature effects from using $N_t=16$
are acceptably small over the whole range of $\mu$ values.

\section{Dependence of the unitarity norm on quark mass and coupling}

It has been observed that the CLE is better behaved if its trajectories remain
near to the $SU(3)$ manifold, i.e. if the unitarity norm remains small. It is
therefore of interest to determine how the unitarity norm depends on the
parameters of the theory.

First we consider how the unitarity norm depends on the quark mass $m$. In the
previous section, we have observed that, for fixed $m$ and $\beta$, the 
local maximum for small $\mu$ occurs at $\mu=0$. The global maximum occurs
at saturation. Since this maximum is the value for the pure gauge theory at
this $\beta$, it does not depend on $m$. Therefore we shall determine the $m$
dependence of the unitarity norm at $\mu=0$, which is thus relevant for small
$\mu$.

Since the unitarity norm at $\mu=0$ appears to decrease with increasing $\beta$
and we shall be interested in $\beta \geq 5.6$, we study the $m$ dependence
with $\beta$ fixed at $\beta=5.6$. We perform CLE simulations with 
$0.01 \le m \le \infty$. For $m=0.01$ we use a $16^4$ lattice. For $m=0.025$
we use both $16^4$ and $12^4$ lattices, while for $m=0.05$, $m=0.1$, $m=0.25$,
$m=0.5$ and $m=\infty$ we use $12^4$ lattices. Except for $m=0.025$ and
$m=\infty$ on $12^4$ lattices, each run uses $3 \times 10^6$ updates.

\begin{figure}[htb]
\epsfxsize=4in
\centerline{\epsffile{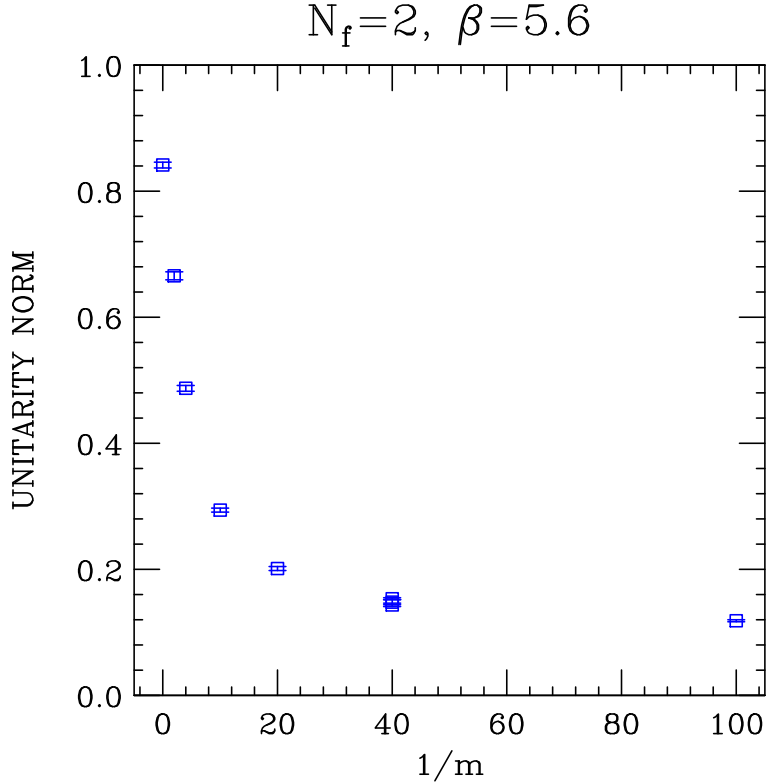}}
\caption{Unitarity norm as a function of inverse quark mass at $\beta=5.6$}
\label{fig:unorm_b56}
\end{figure}

In figure~\ref{fig:unorm_b56} we plot the unitarity norm versus $1/m$ at 
$\beta=5.6$. We note that it decreases by almost an order of magnitude as $m$
is decreased from infinity to $0.01$.

For a given $\beta$, the unitarity norm has its maximum for $\mu$ at
saturation, for which it is mass independent. Hence we choose to perform a CLE
at saturation for each $\beta$, that is we simulate the pure gauge theory for
that $\beta$, knowing that this will give the upper bound to the unitarity norm
for that $\beta$. Moreover, since there are no quarks, we do not have to worry
that we should really change $m$ when we change $\beta$ so as to keep on a line
of constant physics. Since there are no quarks, these simulations are fast,
and for given $\beta$ one can use a much smaller lattice than for the same
$\beta$ with light quarks. Without quarks, the drift term is holomorphic in
the gauge fields, so the CLE should produce results correct to order $dt^2$,
provided the fields evolve on a compact domain.

\begin{figure}[htb]
\epsfxsize=4in
\centerline{\epsffile{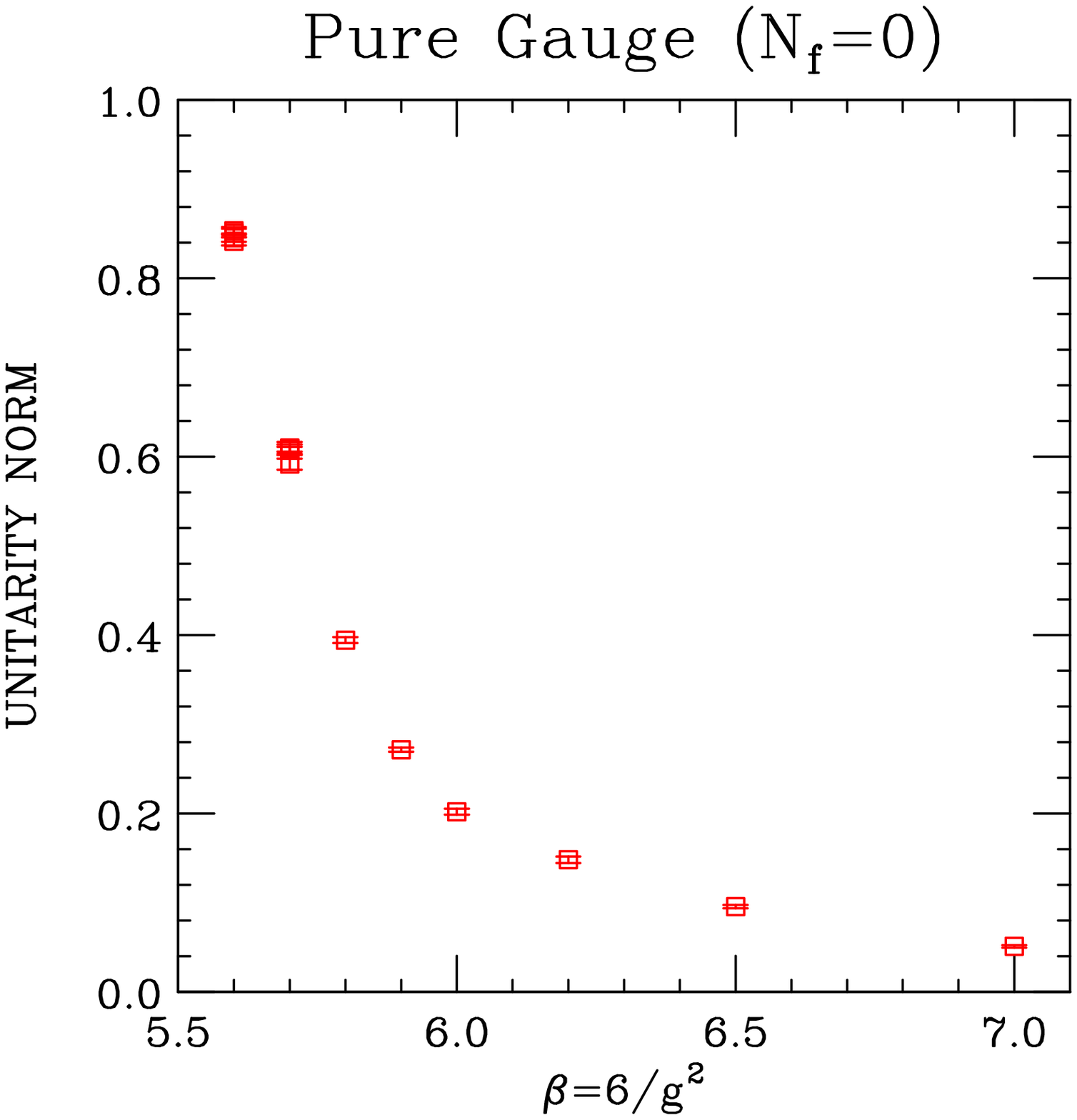}}
\caption{Unitarity norm for pure $SU(3)$ gauge theory as a function of $\beta$.}
\label{fig:unorm_N_f.eq.0}
\end{figure}

We perform CLE simulations for a selection of $\beta$s in the range 
$5.6 \leq \beta \le 7.0$. $\beta=5.6$ and $\beta=5.7$ simulations were
performed on $12^4$ lattices, $\beta=5.8$, $\beta=5.9$, $\beta=6.0$ and
$\beta=6.2$ simulations were performed on $16^4$ lattices, $\beta=6.5$
simulations were performed on a $24^4$ lattice and $\beta=7.0$ simulations
were performed on a $32^4$ lattice. First we find that at $\beta=5.7$, if we
start on the $SU(3)$ manifold, the CLE trajectory stays on the $SU(3)$
manifold for at least $10^7$ updates. However, if we start slightly off this
manifold, the system evolves away from the $SU(3)$ manifold to a region where
the unitarity norm fluctuates around a stable, non-zero value. This value
appears to be independent of how far the starting point is from the $SU(3)$
manifold. We assume similar behaviour for $\beta > 5.7$ and start the
simulations at larger $\beta$s away from the $SU(3)$ manifold. At each
$\beta$, we have at least 1 run with a non-$SU(3)$ start of $5 \times 10^6$
updates or more. The measured unitarity norms are plotted as a function of
$\beta$ in figure~\ref{fig:unorm_N_f.eq.0}. These decrease as $\beta$
increases, that is as the coupling $g$ decreases. In fact over the range of
$\beta$s considered, this norm decreases by more than an order of magnitude.

\begin{figure}[htb]
\epsfxsize=4in
\centerline{\epsffile{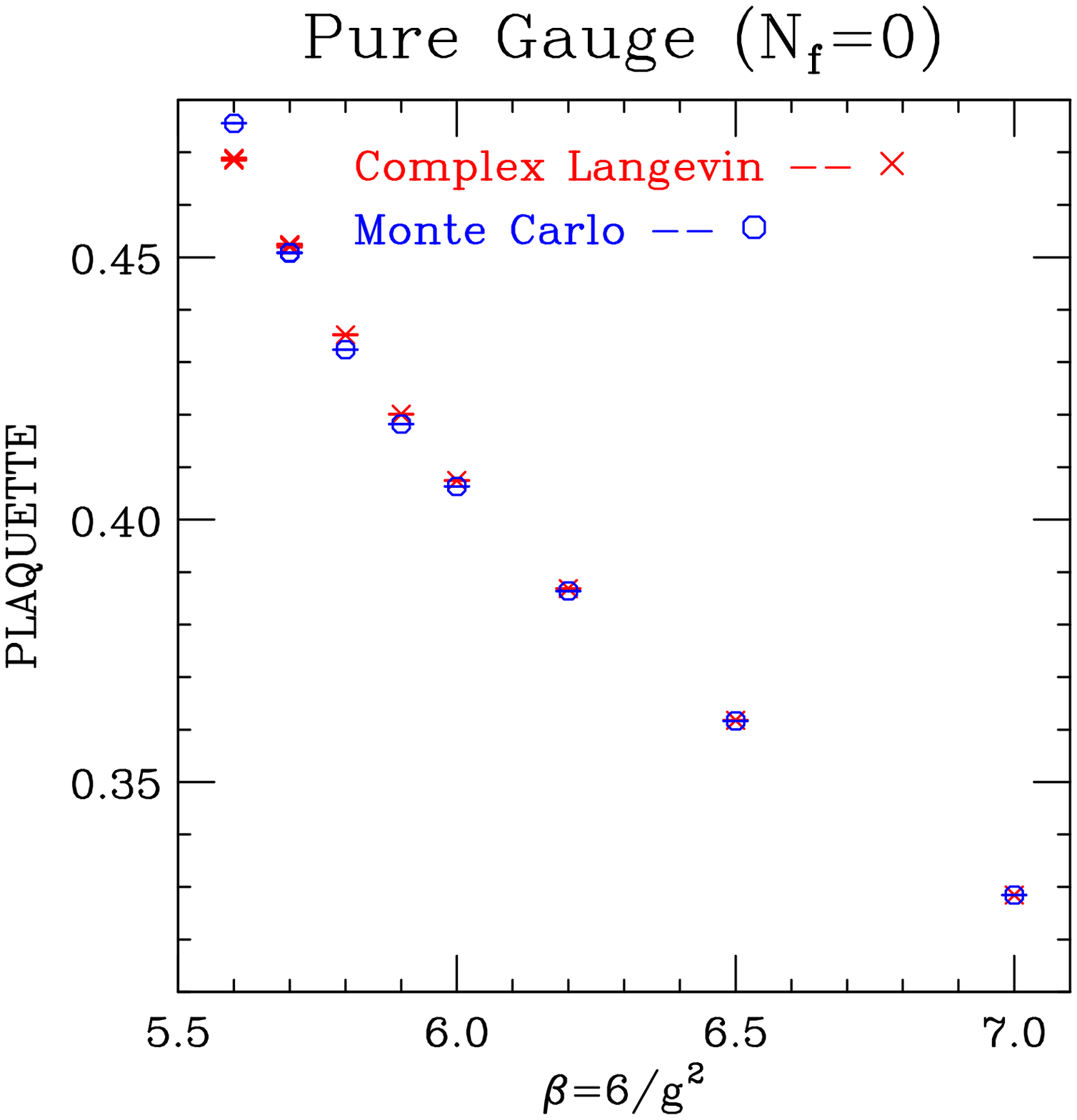}}
\caption{Average plaquette for pure $SU(3)$ gauge theory as a function of 
$\beta$.}
\label{fig:apq_N_f.eq.0}
\end{figure}

In figure~\ref{fig:apq_N_f.eq.0} we plot the average plaquette from our CLE
simulations and compare it to the `exact' value from a Monte-Carlo simulation.
Except for $\beta=5.6$ these values are close and get closer as $\beta$ is 
increased. This difference for the lower $\beta$s is too large to be due to
statistical errors or the inexact nature of Langevin simulations. (This is
surprising since the drift term is holomorphic in the fields, and the region
spanned by these CLE simulations appears to be bounded.) Hence it is a
systematic of the CLE, which must be due to the distribution of the plaquette
values not falling off fast enough at the boundaries or non-ergodicity which
could indicate that there are other regions of the $SL(3,C)^{4 V}$ space which
are not accessible to these simulations. For $\beta=5.6$ where this systematic
error is unacceptably large, we have observed one large excursions in some of
our runs. At all $\beta$s the distributions of plaquette values do have `tails'
or `skirts' as do those of the unitarity norms.

Because of the size of the systematic errors for pure gauge theory at
$\beta=5.6$, we have checked to see if any of this can be due to inadequate
gauge-cooling by running simulations with 5, 7 and 100 cooling steps with
$dt=0.01$. The plaquette values from all 3 are in good agreement, as are the
unitarity norms. We also performed a simulation with 100 cooling steps per
update but with $dt=0.005$, and find good agreement with the plaquettes and
unitarity norms from our $dt=0.01$ simulations. For $\beta=5.7$ we have
performed simulations with 5, 6, 7, 8 and 10 gauge-cooling steps, with
different starting configurations. For the case with 5 gauge-cooling steps per
update, where we used an ordered start, the system stayed in $SU(3)$ (RLE) and
the plaquette was very close to that from the (exact) Monte Carlo value. For
the other 4 choices, the plaquettes were in agreement and only slightly above
the exact value. The agreement between the CLE and exact simulations improves
with increasing $\beta$.

\section{Summary, discussion and conclusions}

We have performed CLE lattice simulations of lattice QCD at zero temperature
and finite $\mu$ at $\beta=5.6$, $m=0.025$ and at $\beta=5.7$, $m=0.025$.
Neither $\beta$ shows the expected physics in the transition region. Whereas
one expects that the transition from hadronic to nuclear matter should occur
at $\mu \approx m_N/3$, these simulations show transitions for $\mu < m_\pi/2$.
For the weaker coupling ($\beta=5.7$) for $\mu$ close to zero, the CLE
produces values of the observables which are close to the correct results, and
considerably better than those for $\beta=5.6$. For $\mu$ large enough to
produce saturation, where all available fermion states are filled, both
$\beta$s indicate that the quarks have decoupled leaving us with a pure
$SU(3)$ gauge theory. In both cases the plaquette observable agrees with that
from a CLE simulation of the pure gauge theory. This value is much closer to
the exact result for the weaker coupling. For $\beta=5.7$ at very small $\mu$
and for a significant range of $\mu$s above $m_N/3$, there is good
agreement between these CLE simulations and the exact (RHMC) simulations of
the phase-quenched theory. 
%For $\mu \ge 0.5$, we argue that the $\beta=5.7$
%observables given by the CLE are probably close to correct.

For $\mu > m_N/3$ we have not seen any sign of new exotic phases, such as a
colour-superconducting phase. Nor is there any indication of a transition to
quark-matter below saturation. There is also no indication of a difference 
between the full theory and its phase-quenched approximation for $\mu \ge 0.5$
while not so high that saturation is a dominant influence, except that the
Wilson Line indicates that the fermion determinant is complex for the full
theory. The apparently gaussian distribution of chiral condensate measurements
(and of quark-number density measurements) suggests that the CLE should be
reliable in this domain. A difference between the full theory and the
phase-quenched theory might be expected because of the existence of a
pion-like superfluid phase in the latter. If the full theory does have local
observables identical to the phase-quenched theory in this region, and this is
not an artifact of the CLE, it might be possible to check this by reweighting
from the phase-quenched theory to the full theory in the high $\mu$ domain.

Because there is some indication that the CLE is more likely to produce correct
results if the trajectories stay close to the $SU(3)$ manifold, we have 
performed CLE simulations over a range of quark masses at a fixed coupling, and
over a range of couplings at infinite quark mass (pure $SU(3)$ gauge theory).
What we find is that the average distance of the trajectories from $SU(3)$ as
determined by the unitarity norm decreases as the coupling and quark mass
are decreased, that is as we approach the continuum limit. This gives us some
hope that the CLE might produce the correct physics in this limit. However, 
random matrix theories suggest that it might produce phase-quenched results.
Since the simulations described in section~3 do not rule out either possibility,
further simulations at weaker couplings, which require larger lattices, are 
needed. Since our conclusion that the CLE fails to observe the transition from
hadronic to nuclear matter at the expected $\mu$ value is based on measurements
of the chiral condensate and quark-number density, both of which are expectation
values of operators with poles at the same place as those in the drift term,
another possibility needs to be considered. That is the possibility that the
simulation is correct and produces the correct values for the expectation
values of non-singular operators but fails for such singular operators. 

Other methods have been suggested to try and obtain correct results from CLE
simulations. One is to add terms to the action which cause the CLE to avoid
the poles, with coefficients, which when taken to zero, yield the original
action \cite{Nagata:2018mkb}. The question then is can these coefficients be
taken small enough to allow them to be continued to zero, without them losing
their effectiveness in avoiding the poles. Preliminary results on very small
lattice look promising, but it remains to be seen if this method will work on
larger lattices.

A second method is based on the observation that one needs to keep the
unitarity norm small for the CLE to work. This is achieved by changing the
dynamics of the CLE by adding a force in the direction of decreasing unitarity
norm to the drift term, with a coefficient which can be made arbitrarily
small \cite{Attanasio:2018rtq}. This additional force should be irrelevant (in
the re-normalization group sense) so that it will vanish as the lattice
spacing goes to zero. It should also vanish when the gauge fields lie on the
$SU(3)$ manifold. Such a force will not be holomorphic in the gauge fields, so
that adding it to the drift term could completely destroy any relationship
between the CLE and the physics contained in the original functional integral,
so careful testing is needed.

Another possible reason for the failure of the CLE has been discussed by Block
and Schenk \cite{Bloch:2017jzi}. Their claim is that part of the problem with
the usual application of the CLE to lattice QCD at finite $\mu$ is the use of
stochastic estimators for the traces of the inverses of the poorly conditioned
Dirac operator. They promote the use of newer methods to replace these
stochastic estimators, which produce better results.

The use of other actions should be investigated. For example, the phase
structure of lattice QCD at zero temperature should be studied as a function
of $\mu$ using the CLE with Wilson fermions, since Wilson fermions preserve
the continuum order of the zeros of the fermion determinant, which is equal to
the number of flavours. For staggered fermions this order is decreased by a
factor of 4 by `taste' breaking, and is only recovered in the continuum limit.

Application of the CLE to finite temperature QCD at finite $\mu$, in particular
with regard to the effect of finite $\mu$ on the transition of hadronic/nuclear
matter to a quark-gluon plasma, should be studied, continuing the pioneering
work of Fodor {\it et al}\cite{Fodor:2015doa}. It has been observed that
the CLE should work better in this case, since finite temperatures move the
zeros of the fermion determinant away from the CLE trajectories. In the case
of 2-flavour staggered quarks, our studies indicate that this is only likely to
work (at least for $m=0.025$) if $\beta=5.6$ lies in the low temperature phase.
This requires $N_t \gtrsim 12$. 

\section*{Acknowledgments}
DKS is supported in part by US Department of Energy contract DE-AC02-06CH11357.
Some of the computing for this research was performed on the Blues and Bebop
clusters at the Laboratory Computing Resource Center at Argonne National 
Laboratory. Part of the computing was performed on the Edison and Cori CRAY
computers at NERSC through an ERCAP allocation. Some of the computing for this
research was provided at the Stampede and Stampede 2 clusters at TACC, the
Bridges cluster at PSC and the Comet cluster at SDSC, through an allocation
provided through XSEDE. DKS would like to thank the CSSM and CoEPP at the
University of Adelaide for their hospitality while some of this research was 
being performed.


\begin{thebibliography}{99}

%%% Complex Langevin %%%

%\cite{Parisi:1984cs}
\bibitem{Parisi:1984cs} 
  G.~Parisi,
  %``On Complex Probabilities,''
  Phys.\ Lett.\ B {\bf 131}, 393 (1983).
  %%CITATION = PHLTA,B131,393;%%
  %148 citations counted in INSPIRE as of 15 Oct 2015

%\cite{Klauder:1983nn}
\bibitem{Klauder:1983nn} 
  J.~R.~Klauder,
  %``Stochastic Quantization,''
  Acta Phys.\ Austriaca Suppl.\  {\bf 25}, 251 (1983).
  %%CITATION = APAUA,25,251;%%
  %33 citations counted in INSPIRE as of 15 Oct 2015

%\cite{Klauder:1983zm}
\bibitem{Klauder:1983zm} 
  J.~R.~Klauder,
  %``A Langevin Approach to Fermion and Quantum Spin Correlation Functions,''
  J.\ Phys.\ A {\bf 16}, L317 (1983).
  %%CITATION = JPAGA,A16,L317;%%
  %41 citations counted in INSPIRE as of 15 Oct 2015

%\cite{Klauder:1983sp}
\bibitem{Klauder:1983sp} 
  J.~R.~Klauder,
  %``Coherent State Langevin Equations for Canonical Quantum Systems With 
  %  Applications to the Quantized Hall Effect,''
  Phys.\ Rev.\ A {\bf 29}, 2036 (1984).
  %%CITATION = PHRVA,A29,2036;%%
  %72 citations counted in INSPIRE as of 15 Oct 2015

%%% Conditions for correct results %%%

%\cite{Aarts:2009uq}
\bibitem{Aarts:2009uq} 
  G.~Aarts, E.~Seiler and I.~O.~Stamatescu,
  %``The Complex Langevin method: When can it be trusted?,''
  Phys.\ Rev.\ D {\bf 81}, 054508 (2010)
  doi:10.1103/PhysRevD.81.054508
  [arXiv:0912.3360 [hep-lat]].
  %%CITATION = doi:10.1103/PhysRevD.81.054508;%%
  %89 citations counted in INSPIRE as of 21 Oct 2016

%\cite{Aarts:2011ax}
\bibitem{Aarts:2011ax} 
  G.~Aarts, F.~A.~James, E.~Seiler and I.~O.~Stamatescu,
  %``Complex Langevin: Etiology and Diagnostics of its Main Problem,''
  Eur.\ Phys.\ J.\ C {\bf 71}, 1756 (2011)
  doi:10.1140/epjc/s10052-011-1756-5
  [arXiv:1101.3270 [hep-lat]].
  %%CITATION = doi:10.1140/epjc/s10052-011-1756-5;%%
  %73 citations counted in INSPIRE as of 21 Oct 2016

%\cite{Nagata:2015uga}
\bibitem{Nagata:2015uga} 
  K.~Nagata, J.~Nishimura and S.~Shimasaki,
  %``Justification of the complex Langevin method with the gauge cooling 
  %procedure,''
  PTEP {\bf 2016}, no. 1, 013B01 (2016)
  doi:10.1093/ptep/ptv173
  [arXiv:1508.02377 [hep-lat]].
  %%CITATION = doi:10.1093/ptep/ptv173;%%
  %14 citations counted in INSPIRE as of 22 Oct 2016

%\cite{Nishimura:2015pba}
\bibitem{Nishimura:2015pba} 
  J.~Nishimura and S.~Shimasaki,
  %``New Insights into the Problem with a Singular Drift Term in the Complex 
  %Langevin Method,''
  Phys.\ Rev.\ D {\bf 92}, no. 1, 011501 (2015)
  doi:10.1103/PhysRevD.92.011501
  [arXiv:1504.08359 [hep-lat]].
  %%CITATION = doi:10.1103/PhysRevD.92.011501;%%
  %27 citations counted in INSPIRE as of 22 Oct 2016

%\cite{Nagata:2016vkn}
\bibitem{Nagata:2016vkn} 
  K.~Nagata, J.~Nishimura and S.~Shimasaki,
  %``Argument for justification of the complex Langevin method and the condition
  %for correct convergence,''
  Phys.\ Rev.\ D {\bf 94}, no. 11, 114515 (2016)
  doi:10.1103/PhysRevD.94.114515
  [arXiv:1606.07627 [hep-lat]].
  %%CITATION = doi:10.1103/PhysRevD.94.114515;%%
  %22 citations counted in INSPIRE as of 09 Oct 2017

%\cite{Aarts:2017vrv}
\bibitem{Aarts:2017vrv} 
  G.~Aarts, E.~Seiler, D.~Sexty and I.~O.~Stamatescu,
  %``Complex Langevin dynamics and zeroes of the fermion determinant,''
  JHEP {\bf 1705}, 044 (2017)
  doi:10.1007/JHEP05(2017)044
  [arXiv:1701.02322 [hep-lat]].
  %%CITATION = doi:10.1007/JHEP05(2017)044;%%
  %8 citations counted in INSPIRE as of 09 Oct 2017

%\cite{Seiler:2017wvd}
\bibitem{Seiler:2017wvd} 
  E.~Seiler,
  %``Status of Complex Langevin,''
  EPJ Web Conf.\  {\bf 175}, 01019 (2018)
  doi:10.1051/epjconf/201817501019
  [arXiv:1708.08254 [hep-lat]].
  %%CITATION = doi:10.1051/epjconf/201817501019;%%
  %15 citations counted in INSPIRE as of 12 Feb 2019

%\cite{Aarts:2017hqp}
\bibitem{Aarts:2017hqp} 
  G.~Aarts, K.~Boguslavski, M.~Scherzer, E.~Seiler, D.~Sexty and 
  I.~O.~Stamatescu,
  %``Getting even with CLE,''
  EPJ Web Conf.\  {\bf 175}, 14007 (2018)
  doi:10.1051/epjconf/201817514007
  [arXiv:1710.05699 [hep-lat]].
  %%CITATION = doi:10.1051/epjconf/201817514007;%%

%\cite{Nagata:2018net}
\bibitem{Nagata:2018net} 
  K.~Nagata, J.~Nishimura and S.~Shimasaki,
  %``Testing the criterion for correct convergence in the complex Langevin method,''
  JHEP {\bf 1805}, 004 (2018)
  doi:10.1007/JHEP05(2018)004
  [arXiv:1802.01876 [hep-lat]].
  %%CITATION = doi:10.1007/JHEP05(2018)004;%%
  %3 citations counted in INSPIRE as of 11 Feb 2019

%%% Seminal paper on gauge cooling %%%

%\cite{Seiler:2012wz}
\bibitem{Seiler:2012wz} 
  E.~Seiler, D.~Sexty and I.~O.~Stamatescu,
  %``Gauge cooling in complex Langevin for QCD with heavy quarks,''
  Phys.\ Lett.\ B {\bf 723}, 213 (2013)
  [arXiv:1211.3709 [hep-lat]].
  %%CITATION = ARXIV:1211.3709;%%
  %45 citations counted in INSPIRE as of 08 Oct 2015


%%% QCD with heavy quarks %%%

%\cite{Aarts:2008rr}
\bibitem{Aarts:2008rr} 
  G.~Aarts and I.~O.~Stamatescu,
  %``Stochastic quantization at finite chemical potential,''
  JHEP {\bf 0809}, 018 (2008)
  doi:10.1088/1126-6708/2008/09/018
  [arXiv:0807.1597 [hep-lat]].
  %%CITATION = doi:10.1088/1126-6708/2008/09/018;%%
  %93 citations counted in INSPIRE as of 20 Oct 2016

%\cite{Aarts:2013uxa}
\bibitem{Aarts:2013uxa} 
  G.~Aarts, L.~Bongiovanni, E.~Seiler, D.~Sexty and I.~O.~Stamatescu,
  %``Controlling complex Langevin dynamics at finite density,''
  Eur.\ Phys.\ J.\ A {\bf 49}, 89 (2013)
  doi:10.1140/epja/i2013-13089-4
  [arXiv:1303.6425 [hep-lat]].
  %%CITATION = doi:10.1140/epja/i2013-13089-4;%%
  %70 citations counted in INSPIRE as of 20 Oct 2016

%\cite{Aarts:2014bwa}
\bibitem{Aarts:2014bwa} 
  G.~Aarts, E.~Seiler, D.~Sexty and I.~O.~Stamatescu,
  %``Simulating QCD at nonzero baryon density to all orders in the hopping 
  %parameter expansion,''
  Phys.\ Rev.\ D {\bf 90}, no. 11, 114505 (2014)
  doi:10.1103/PhysRevD.90.114505
  [arXiv:1408.3770 [hep-lat]].
  %%CITATION = doi:10.1103/PhysRevD.90.114505;%%
  %35 citations counted in INSPIRE as of 20 Oct 2016

%\cite{Aarts:2016qrv}
\bibitem{Aarts:2016qrv} 
  G.~Aarts, F.~Attanasio, B.~J\"{a}ger and D.~Sexty,
  %``The QCD phase diagram in the limit of heavy quarks using complex Langevin 
  %dynamics,''
  JHEP {\bf 1609}, 087 (2016)
  doi:10.1007/JHEP09(2016)087
  [arXiv:1606.05561 [hep-lat]].
  %%CITATION = doi:10.1007/JHEP09(2016)087;%%
  %4 citations counted in INSPIRE as of 20 Oct 2016

%\cite{Langelage:2014vpa}
\bibitem{Langelage:2014vpa} 
  J.~Langelage, M.~Neuman and O.~Philipsen,
  %``Heavy dense QCD and nuclear matter from an effective lattice theory,''
  JHEP {\bf 1409}, 131 (2014)
  doi:10.1007/JHEP09(2014)131
  [arXiv:1403.4162 [hep-lat]].
  %%CITATION = doi:10.1007/JHEP09(2014)131;%%
  %22 citations counted in INSPIRE as of 21 Oct 2016

%\cite{Rindlisbacher:2015pea}
\bibitem{Rindlisbacher:2015pea} 
  T.~Rindlisbacher and P.~de Forcrand,
  %``Two-flavor lattice QCD with a finite density of heavy quarks: heavy-dense 
  % limit and “particle-hole” symmetry,''
  JHEP {\bf 1602}, 051 (2016)
  doi:10.1007/JHEP02(2016)051
  [arXiv:1509.00087 [hep-lat]].
  %%CITATION = doi:10.1007/JHEP02(2016)051;%%
  %15 citations counted in INSPIRE as of 23 Jul 2019

%%% Full QCD at smaller masses %%%

%\cite{Sexty:2013ica}
\bibitem{Sexty:2013ica} 
  D.~Sexty,
  %``Simulating full QCD at nonzero density using the complex Langevin 
  %  equation,''
  Phys.\ Lett.\ B {\bf 729}, 108 (2014)
  [arXiv:1307.7748 [hep-lat]].
  %%CITATION = ARXIV:1307.7748;%%
  %54 citations counted in INSPIRE as of 09 Oct 2015

%\cite{Fodor:2015doa}
\bibitem{Fodor:2015doa} 
  Z.~Fodor, S.~D.~Katz, D.~Sexty and C.~T\"{o}r\"{o}k,
  %``Complex Langevin dynamics for dynamical QCD at nonzero chemical 
  %potential: A comparison with multiparameter reweighting,''
  Phys.\ Rev.\ D {\bf 92}, no. 9, 094516 (2015)
  doi:10.1103/PhysRevD.92.094516
  [arXiv:1508.05260 [hep-lat]].
  %%CITATION = doi:10.1103/PhysRevD.92.094516;%%
  %15 citations counted in INSPIRE as of 21 Oct 2016 

%\cite{Nagata:2016mmh}
\bibitem{Nagata:2016mmh} 
  K.~Nagata, H.~Matsufuru, J.~Nishimura and S.~Shimasaki,
  %``Gauge cooling for the singular-drift problem in the complex Langevin method
  %--- an application to finite density QCD,''
  PoS LATTICE {\bf 2016}, 067 (2016)
  [arXiv:1611.08077 [hep-lat]].
  %%CITATION = ARXIV:1611.08077;%%
  %4 citations counted in INSPIRE as of 10 Oct 2017

%\cite{Tsutsui:2018jva}
\bibitem{Tsutsui:2018jva} 
  S.~Tsutsui, Y.~Ito, H.~Matsufuru, J.~Nishimura, S.~Shimasaki and A.~Tsuchiya,
  %``Can the complex Langevin method see the deconfinement phase transition in 
  % QCD at finite density?,''
  arXiv:1811.07647 [hep-lat].
  %%CITATION = ARXIV:1811.07647;%%

%\cite{Scherzer:2018udt}
\bibitem{Scherzer:2018udt} 
  M.~Scherzer, E.~Seiler, D.~Sexty and I.~O.~Stamatescu,
  %``Complex Langevin: Boundary terms and application to QCD,''
  arXiv:1810.09713 [hep-lat].
  %%CITATION = ARXIV:1810.09713;%%
  %1 citations counted in INSPIRE as of 11 Feb 2019

%\cite{Ito:2018jpo}
\bibitem{Ito:2018jpo} 
  Y.~Ito, H.~Matsufuru, J.~Nishimura, S.~Shimasaki, A.~Tsuchiya and S.~Tsutsui,
  %``Exploring the phase diagram of finite density QCD at low temperature by 
  % the complex Langevin method,''
  arXiv:1811.12688 [hep-lat].
  %%CITATION = ARXIV:1811.12688;%%

%%% RMT relating CLE for QCD at finite mu and phase-quenched theory

%\cite{Mollgaard:2013qra}
\bibitem{Mollgaard:2013qra} 
  A.~Mollgaard and K.~Splittorff,
  %``Complex Langevin Dynamics for chiral Random Matrix Theory,''
  Phys.\ Rev.\ D {\bf 88}, no. 11, 116007 (2013)
  doi:10.1103/PhysRevD.88.116007
  [arXiv:1309.4335 [hep-lat]].
  %%CITATION = doi:10.1103/PhysRevD.88.116007;%%
  %76 citations counted in INSPIRE as of 22 Jul 2019

%\cite{Osborn:2004rf}
\bibitem{Osborn:2004rf} 
  J.~C.~Osborn,
  %``Universal results from an alternate random matrix model for QCD with a 
  % baryon chemical potential,''
  Phys.\ Rev.\ Lett.\  {\bf 93}, 222001 (2004)
  doi:10.1103/PhysRevLett.93.222001
  [hep-th/0403131].
  %%CITATION = doi:10.1103/PhysRevLett.93.222001;%%
  %116 citations counted in INSPIRE as of 22 Jul 2019

%\cite{Bloch:2012bh}
\bibitem{Bloch:2012bh} 
  J.~Bloch, F.~Bruckmann, M.~Kieburg, K.~Splittorff and J.~J.~M.~Verbaarschot,
  %``Subsets of configurations and canonical partition functions,''
  Phys.\ Rev.\ D {\bf 87}, no. 3, 034510 (2013)
  doi:10.1103/PhysRevD.87.034510
  [arXiv:1211.3990 [hep-lat]].
  %%CITATION = doi:10.1103/PhysRevD.87.034510;%%
  %18 citations counted in INSPIRE as of 22 Jul 2019

%\cite{Mollgaard:2014mga}
\bibitem{Mollgaard:2014mga} 
  A.~Mollgaard and K.~Splittorff,
  %``Full simulation of chiral random matrix theory at nonzero chemical potential by complex Langevin,''
  Phys.\ Rev.\ D {\bf 91}, no. 3, 036007 (2015)
  doi:10.1103/PhysRevD.91.036007
  [arXiv:1412.2729 [hep-lat]].
  %%CITATION = doi:10.1103/PhysRevD.91.036007;%%
  %34 citations counted in INSPIRE as of 22 Jul 2019

%\cite{Bloch:2017sex}
\bibitem{Bloch:2017sex} 
  J.~Bloch, J.~Glesaaen, J.~J.~M.~Verbaarschot and S.~Zafeiropoulos,
  %``Complex Langevin Simulation of a Random Matrix Model at Nonzero Chemical 
  %Potential,''
  JHEP {\bf 1803}, 015 (2018)
  doi:10.1007/JHEP03(2018)015
  [arXiv:1712.07514 [hep-lat]].
  %%CITATION = doi:10.1007/JHEP03(2018)015;%%
  %5 citations counted in INSPIRE as of 14 Oct 2018

%\cite{Stephanov:1996ki}
\bibitem{Stephanov:1996ki} 
  M.~A.~Stephanov,
  %``Random matrix model of QCD at finite density and the nature of the 
  %  quenched limit,''
  Phys.\ Rev.\ Lett.\  {\bf 76}, 4472 (1996)
  doi:10.1103/PhysRevLett.76.4472
  [hep-lat/9604003].
  %%CITATION = doi:10.1103/PhysRevLett.76.4472;%%
  %295 citations counted in INSPIRE as of 22 Jul 2019

%\cite{Nagata:2016alq}
\bibitem{Nagata:2016alq} 
  K.~Nagata, J.~Nishimura and S.~Shimasaki,
  %``Gauge cooling for the singular-drift problem in the complex Langevin method
  % - a test in Random Matrix Theory for finite density QCD,''
  JHEP {\bf 1607}, 073 (2016)
  doi:10.1007/JHEP07(2016)073
  [arXiv:1604.07717 [hep-lat]].
  %%CITATION = doi:10.1007/JHEP07(2016)073;%%
  %12 citations counted in INSPIRE as of 10 Oct 2017


%%% Our Lattice 2018 contribution %%%

%\cite{Sinclair:2018rbk}
\bibitem{Sinclair:2018rbk} 
  D.~K.~Sinclair and J.~B.~Kogut,
  %``Complex Langevin for Lattice QCD,''
  arXiv:1810.11880 [hep-lat].
  %%CITATION = ARXIV:1810.11880;%%
  %1 citations counted in INSPIRE as of 11 Feb 2019

%%% Partial second order formalism for Langevin equation %%%

%\cite{Ukawa:1985hr}
\bibitem{Ukawa:1985hr}
  A.~Ukawa and M.~Fukugita,
  %``Langevin Simulation Including Dynamical Quark Loops,''
  Phys.\ Rev.\ Lett.\  {\bf 55} (1985) 1854.
  %%CITATION = PRLTA,55,1854;%%
  %99 citations counted in INSPIRE as of 09 Oct 2015

%\cite{Fukugita:1986tg}
\bibitem{Fukugita:1986tg}
  M.~Fukugita, Y.~Oyanagi and A.~Ukawa,
  %``Langevin Simulation of the Full {QCD} Hadron Mass Spectrum on a Lattice,''
  Phys.\ Rev.\ D {\bf 36} (1987) 824.
  %%CITATION = PHRVA,D36,824;%%
  %55 citations counted in INSPIRE as of 09 Oct 2015

%\cite{Fukugita:1988qs}
\bibitem{Fukugita:1988qs} 
  M.~Fukugita and A.~Ukawa,
  %``Finite Temperature Phase Transitions in Lattice {QCD} With Langevin 
  %Simulation,''
  Phys.\ Rev.\ D {\bf 38}, 1971 (1988).
  %%CITATION = PHRVA,D38,1971;%%
  %23 citations counted in INSPIRE as of 09 Oct 2015

%%% Finite isospin density paper %%%

%\cite{Kogut:2002zg}
\bibitem{Kogut:2002zg} 
  J.~B.~Kogut and D.~K.~Sinclair,
  %``Lattice QCD at finite isospin density at zero and finite temperature,''
  Phys.\ Rev.\ D {\bf 66}, 034505 (2002)
  doi:10.1103/PhysRevD.66.034505
  [hep-lat/0202028].
  %%CITATION = doi:10.1103/PhysRevD.66.034505;%%
  %193 citations counted in INSPIRE as of 03 Mar 2019

%%% HEMCGC collaboration simulations at $\beta=5.6$ %%%

%\cite{Bitar:1990cb}
\bibitem{Bitar:1990cb} 
  K.~M.~Bitar {\it et al.},
  %``Hadron spectrum in QCD at 6/g**2 = 5.6,''
  Phys.\ Rev.\ D {\bf 42}, 3794 (1990).
  %%CITATION = PHRVA,D42,3794;%%
  %51 citations counted in INSPIRE as of 09 Oct 2015

%%% Pion and nucleon masses at beta=5.7, m=0.025

%\cite{Brown:1991qw}
\bibitem{Brown:1991qw} 
  F.~R.~Brown, F.~P.~Butler, H.~Chen, N.~H.~Christ, Z.~h.~Dong, W.~Schaffer, 
  L.~I.~Unger and A.~Vaccarino,
  %``Hadron masses in QCD with two flavors of dynamical fermions at Beta = 5.7,
  %''
  Phys.\ Rev.\ Lett.\  {\bf 67}, 1062 (1991).
  doi:10.1103/PhysRevLett.67.1062
  %%CITATION = doi:10.1103/PhysRevLett.67.1062;%%
  %69 citations counted in INSPIRE as of 11 Oct 2017

%\cite{Schaffer:1992rq}
\bibitem{Schaffer:1992rq} 
  W.~Schaffer,
  %``Hadron masses on a 32**4 lattice at beta = 5.7,''
  Nucl.\ Phys.\ Proc.\ Suppl.\  {\bf 30}, 405 (1993).
  doi:10.1016/0920-5632(93)90238-2
  %%CITATION = doi:10.1016/0920-5632(93)90238-2;%%
  %3 citations counted in INSPIRE as of 11 Oct 2017

%%%% determinant Polyakov Loop correlations %%%

%\cite{deForcrand:1999cy}
\bibitem{deForcrand:1999cy} 
  P.~de Forcrand and V.~Laliena,
  %``The Role of the Polyakov loop in finite density QCD,''
  Phys.\ Rev.\ D {\bf 61}, 034502 (2000)
  doi:10.1103/PhysRevD.61.034502
  [hep-lat/9907004].
  %%CITATION = doi:10.1103/PhysRevD.61.034502;%%
  %24 citations counted in INSPIRE as of 31 Jul 2019

%%%% Deformation Technique %%%

%\cite{Nagata:2018mkb}
\bibitem{Nagata:2018mkb} 
  K.~Nagata, J.~Nishimura and S.~Shimasaki,
  %``Complex Langevin calculations in finite density QCD at large μ/T with the deformation technique,''
  Phys.\ Rev.\ D {\bf 98}, no. 11, 114513 (2018)
  doi:10.1103/PhysRevD.98.114513
  [arXiv:1805.03964 [hep-lat]].
  %%CITATION = doi:10.1103/PhysRevD.98.114513;%%
  %4 citations counted in INSPIRE as of 12 Feb 2019

%%%% Dynamic Stabilization %%%

%\cite{Attanasio:2018rtq}
\bibitem{Attanasio:2018rtq} 
  F.~Attanasio and B.~J\"{a}ger,
  %``Dynamical stabilisation of complex Langevin simulations of QCD,''
  Eur.\ Phys.\ J.\ C {\bf 79}, no. 1, 16 (2019)
  doi:10.1140/epjc/s10052-018-6512-7
  [arXiv:1808.04400 [hep-lat]].
  %%CITATION = doi:10.1140/epjc/s10052-018-6512-7;%%
  %5 citations counted in INSPIRE as of 12 Feb 2019

%%%% Replacing Stochastic Estimators %%%%%

%\cite{Bloch:2017jzi}
\bibitem{Bloch:2017jzi} 
  J.~Bloch and O.~Schenk,
  %``Selected inversion as key to a stable Langevin evolution across the QCD phase boundary,''
  EPJ Web Conf.\  {\bf 175}, 07003 (2018)
  doi:10.1051/epjconf/201817507003
  [arXiv:1707.08874 [hep-lat]].
  %%CITATION = doi:10.1051/epjconf/201817507003;%%
  %6 citations counted in INSPIRE as of 01 Aug 2019

\end{thebibliography}
\end{document}